\newif\ifShowKeys
\ShowKeystrue
\ShowKeysfalse

\documentclass[11pt]{article}
	\pdfoutput=1
\usepackage[usenames,dvipsnames]{xcolor}
\usepackage[setpagesize=false,pagebackref=false, linktocpage, bookmarksopen=true, colorlinks=true, linkcolor=Maroon,citecolor=Maroon,urlcolor=Maroon]{hyperref}
\usepackage[parsep]{collref}

\ifShowKeys \usepackage[notcite]{showkeys} \fi

\usepackage{amsmath, amssymb,amsthm}
\numberwithin{equation}{section}

\usepackage{bm,environ,mathrsfs,array,arydshln}
\usepackage{booktabs,float,slashed,hyperref}
\usepackage[mathcal]{euscript}
\usepackage{tensor} 						
\usepackage{mathabx}
\usepackage[vcentermath]{youngtab}
\usepackage{xcolor}

\usepackage{aurical}
\usepackage[T1]{fontenc}
\usepackage[nodayofweek]{date time}
\usepackage{graphicx,epsfig,epic}
\usepackage{tikz} 
\usetikzlibrary{arrows,decorations.pathreplacing,decorations.markings,snakes}
\usetikzlibrary{cd}

\usepackage{framed}						
\definecolor{shadecolor}{rgb}{0.9996078, 0.984314, 0.960784}

\usepackage{comment}

\allowdisplaybreaks

\catcode`,\active

\catcode`\,12

\definecolor{myred}{RGB}{233, 33, 45}

\newcommand{\bs}{\begin{shaded}}
\newcommand{\es}{\end{shaded}\noindent}
\def\ba#1\ea{\begin{align}#1\end{align}}		
\newcommand{\be}{\begin{equation}}
\newcommand{\ee}{\end{equation}}
\newcommand{\bea}{\begin{equation} \begin{aligned}} 
\newcommand{\eea}{\end{aligned} \end{equation}}
\newcommand{\mc}{\mathcal }

\newcommand{\la}{\label}
\newcommand{\eps}{\varepsilon}

\newcommand{\lp}{\notag \\ & }

\newcommand{\wt}{\widetilde}

\newcommand{\cf}{\textit{cf.} }
\newcommand{\ie}{\textit{i.e.} }

\newcommand{\N}{\mathcal N}

\renewcommand{\l}{\lambda}

\DeclareMathOperator{\Tr}{Tr}
\DeclareMathOperator{\PE}{PE}

\newcommand{\I}{\mathrm{I}}

\newcommand{\vth}{\vartheta}
\newcommand{\z}{\rho}
\newcommand{\bz}{\bm{z}}
\newcommand{\bg}{\bm{g}}

\newcommand{\wh}{\widehat}

\newcommand{\smb}{\scalebox{0.6}{$\Box$}}

\begin{document}

\begin{titlepage}

\vspace*{15mm}
\begin{center}
{\Large\sc   Leading Giant graviton expansion of }\vskip 9pt
{\Large\sc       Schur  correlators in large representations}

\vspace*{10mm}

{\Large M. Beccaria}

\vspace*{4mm}
	
${}^a$ Universit\`a del Salento, Dipartimento di Matematica e Fisica \textit{Ennio De Giorgi},\\ 
		and INFN - sezione di Lecce, Via Arnesano, I-73100 Lecce, Italy
			\vskip 0.3cm
\vskip 0.2cm {\small E-mail: \texttt{matteo.beccaria@le.infn.it}}
\vspace*{0.8cm}
\end{center}

\begin{abstract}  
We consider 4d $\N=4$ $U(N)$ SYM and the leading giant graviton correction to the Schur defect 2-point 
functions of  $\frac{1}{2}$-BPS Wilson lines in rank-$k$ symmetric and antisymmetric  representations.
We study in particular the large $k$ limit for the symmetric case and the regime $1\ll k \ll N$ in the antisymmetric one.
We present exact results for the correction in agreement with matrix model evaluation at finite $N,k$. 
The Wilson lines in symmetric/antisymmetric representations admit a description in terms of D3$_{k}$ and D5$_{k}$ brane probes
representing a collection of $k$ fundamental strings. In this picture, giant graviton corrections come from fluctuations
of brane probes in presence of a wrapped D3 brane giant graviton. In particular, for the antisymmetric case, our leading correction 
matches  the half-index of the 3d $\mathcal N=4$ Maxwell theory living on the 3d disk which is a part of the giant graviton
divided out by the D5$_{k}$ probe, as recently proposed in arXiv:2404.08302. For the symmetric case at large $k$, we derive an explicit exact 
residue formula for the leading correction.
\end{abstract}
\vskip 0.5cm
	{
	}
\end{titlepage}

\tableofcontents
\vspace{1cm}

\section{Introduction and results}

The superconformal index  \cite{Kinney:2005ej,Romelsberger:2005eg,Romelsberger:2007ec}
of 4d $\N=4$ $U(N)$ SYM at large $N$ and fixed charge is the generating function of BPS  single trace states
and has finite $N$ corrections associated with $U(N)$ trace relations in multi-trace states. \footnote{For the relation between 
large $N$ limit of the full index and black hole physics, see \cite{Agarwal:2020zwm,Murthy:2020rbd}.}
In the dual IIB superstring theory in $AdS_{5}\times S^{5}$, the large $N$ limit reproduces the contributions to the index 
from BPS supergravity states. 
Finite $N$ corrections come from 
wrapped D3 brane configurations with charge $\sim N$ \cite{Imamura:2021ytr,Gaiotto:2021xce,Lee:2022vig} and may 
be generically named (generalized) giant graviton contributions \cite{McGreevy:2000cw,Mikhailov:2000ya}. \footnote{
Alternatively, giant graviton contributions can also be examined in terms of supergravity bubbling solutions \cite{Chang:2024zqi, DeddoLiu}.
} The Schur specialization of the superconformal index \cite{Gadde:2011ik,Gadde:2011uv,Beem:2013sza} is defined as the following trace 
\be
\I^{U(N)}(\eta; q) = \Tr_{\rm BPS}[(-1)^{\rm F}\, q^{H+J+\bar J}\eta^{R}]\, .
\ee
It depends on two fugacities coupled to Cartan generators of the superconformal symmetry group $PSU(2,2|4)$ commuting with the supercharges
entering the BPS condition. In more details, $q$ is the  fugacity coupled to the Hamiltonian $H$ and the two spins $J, \bar J$, while $\eta$ is a flavor
fugacity coupled to an $R$-charge generator. The Schur index has an explicit expression as the singlet projection of many-particle contributions 
built out of the Schur single-particle index $f(\eta; q)$ by usual plethystic \footnote{For a set of fugacities, group matrices, \textit{etc.}, collectively 
denoted by $\bm{x}$ one has $\PE[F(\bm{x})] = \exp\sum_{n\ge 1}\frac{1}{n}F(\bm{x}^{n})$.
}
\bea
\la{1.2}
\I^{U(N)}(\eta; q) & = \int_{U(N)} DU\  \PE[ f(\eta; q)\, \Tr U\, \Tr U^{-1}],\qquad
f(\eta; q) = \frac{(\eta+\eta^{-1})\, q-2q^{2}}{1-q^{2}}\,.
\eea
This matrix integral was computed exactly at finite $N$ in \cite{Bourdier:2015sga,Bourdier:2015wda,Pan:2021mrw,Hatsuda:2022xdv}, and 
for other gauge groups in \cite{Du:2023kfu}. The structure of its large and finite $N$ corrections was first studied in \cite{Arai:2020qaj}
and takes the form 
\be
\la{1.3}
\I^{U(N)}(\eta; q) = \I^{U(\infty)}(\eta;q)\,\bigg[1+\sum_{n\ge 1}q^{nN}\, \wh\I_{n}(\eta; q)\bigg]\, ,
\ee
where $\I^{U(\infty)}(\eta;q)$ is the large $N$  supergravity contribution, while  
terms $\wh\I_{n}(\eta; q)$ come from wrapped $S^{1}\times S^{3}$ D3 branes in $AdS_{5}\times S^{5}$
with  $S^{1}\subset AdS_{5}$ and $S^{3}\subset S^{5}$. \footnote{
For a direct D3 brane fluctuation analysis reproducing the leading correction in (\ref{1.3}), see 
\cite{Beccaria:2023sph,Beccaria:2023cuo,Beccaria:2024vfx,Gautason:2024nru}. Higher giant-graviton contributions
for the $\frac{1}{2}$-BPS  index were analyzed by localization on the brane world volume  \cite{Lee:2023iil,Eleftheriou:2023jxr}.
}
They  may be given by analytic continuation of the $\N=4$ $U(n)$ SYM index  \cite{Arai:2020qaj}, following the approach in 
\cite{Arai:2019wgv,Arai:2019aou,Arai:2020uwd,Fujiwara:2021xgu,Imamura:2021dya,Imamura:2022aua,Fujiwara:2023bdc}
and were computed exactly in \cite{Beccaria:2024szi}.

\paragraph{Schur line correlators and string fluctuations}
 
The Schur index is a supersymmetric partition function on $S^{1}\times S^{3}$ and it is possible to consider 
additional defect 't Hooft or Wilson lines along $S^{1}$ and at positions on a great circle of $S^{3}$ 
\cite{Dimofte:2011py,Gang:2012yr,Cordova:2016uwk,Pan:2019bor}.
The resulting correlation functions are actually topological, \ie do not depend on the line positions.  Wilson lines in 
general representations $\mathsf R_{1}, \mathsf R_{2}, \dots$ corresponds to the following generalization of (\ref{1.2}) \footnote{
In a basis where $U=\text{diag}(e^{iz_{1}}, \dots, e^{iz_{n}})$, the trace $\Tr U$ is the character of the fundamental representation $\Tr U = \chi_{\smb}(z)$
and $\Tr_{\mathsf R}U$ stands similarly for $\chi_{\mathsf R}(z)$.
}
\ba
\la{1.4}
\I^{U(N)}_{\mathsf R_{1}, \mathsf R_{2}, \dots}(\eta; q) & = \int_{U(N)}DU\, \prod_{n\ge 1}\Tr_{\mathsf R_{n}}(U)\, 
\PE[ f(\eta; q)\Tr U\, \Tr U^{-1}]\,,
\ea
 and exact results were obtained in \cite{Drukker:2015spa,Cordova:2016uwk,Neitzke:2017cxz,Hatsuda:2023iwi,Hatsuda:2023imp,Hatsuda:2023iof,Guo:2023mkn}
at fixed or infinite $N$.

The giant graviton expansion of Schur line correlators (\ref{1.4}) was recently considered
in \cite{Imamura:2024lkw,Beccaria:2024oif}. In the simplest case of the defect 2-point function with one line in the fundamental and one in the antifundamental 
$\I^{U(N)}_{\rm F}(\eta; q)\equiv \I^{U(N)}_{\smb, \overline{\smb}}(\eta; q)$, the large $N$ limit has the simple factorized form  \cite{Gang:2012yr}
\be
\I_{\rm F}^{U(\infty)}(\eta; q) = \I_{\rm F1}(\eta; q)\  \I^{U(\infty)}(\eta; q)\,, 
\ee
where the additional factor $\I_{\rm F1}(\eta; q)$ reads
\be
\I_{\rm F1}(\eta; q) = \frac{1}{1-f(\eta; q)} = \PE[f_{\rm F1}(\eta; q)], \qquad f_{\rm F1}(\eta; q) = -q^{2}+(\eta+\eta^{-1})\, q\, .
\ee 
Its gravity interpretation  is expected to involve fluctuations of a fundamental string stretched along $AdS_{2}$ inside $AdS_{5}$ \cite{Rey:1998ik,Maldacena:1998im}.
They fill a short multiplet of $OSp(4^{*}|4)$ with $8_{B}+8_{F}$ states and the three terms in $f_{\rm F1}$  agree with the contributions to the single particle index from  three BPS modes \cite{Gang:2012yr}.
At finite $N$, the leading single giant graviton correction to $\I_{\rm F}^{U(N)}(\eta; q)$ is due to fluctuations of a world-sheet bounded by 
two semi-infinite strings attached to the Wilson lines and ending on the giant graviton, \ie a wrapped D3 brane \cite{Imamura:2024lkw}.
Explicitly, at leading order, one has 
\ba
\la{1.7}
& \frac{\I_{\rm F}^{U(N)}-I_{\rm F1}\,\I^{U(N)}}{\I^{U(\infty)}}  = 
1+\bigg(\mathsf{G}^{+}_{\rm F}(\eta; q)\, \eta^{N}+\mathsf{G}^{-}_{\rm F}(\eta; q)\,\eta^{-N}\bigg)\, q^{N}+\cdots, \quad 
\mathsf{G}^{-}_{\rm F}(\eta; q) = \mathsf{G}^{+}_{\rm F}(\eta^{-1}; q)\,.
\ea
The functions $\mathsf{G}^{\pm}$ factorize  and read \cite{Beccaria:2024oif,Imamura:2024lkw}
(see Appendix \ref{app:special} for special function conventions)
\ba
\la{1.8}
& \mathsf{G}^{\pm}_{\rm F}(\eta; q)  =  \frac{1}{\eta q}\,  \PE[f_{\rm F}(\eta; q) ]\,\,G^{\pm}_{\rm D3}(\eta; q)\,, 
\qquad G^{+}_{\rm D3}(\eta; q) =  -\eta^{2}q\, \frac{(\frac{q}{\eta})_{\infty}^{3}}{\vth(\eta^{2},\frac{q}{\eta})}\, ,
\ea
where  $G^{\pm}_{\rm D3}(\eta; q)$ is the leading giant graviton contribution to the  Schur index without insertions computed from wrapped D3 branes in
 \cite{Arai:2020qaj,Beccaria:2024szi}, and 
\be
\la{1.9}
f_{\rm F}(\eta; q) = 2\eta^{-1}q-2q^{2}\,,
\ee
is the single particle index of fluctuations of the system of two semi-infinite strings. Up to a factor 2 accounting for the two strings, it selects two terms 
out of the three in  $f_{\rm F1}$ since one BPS state is missing due to the boundary condition that they should end on the giant graviton \cite{Imamura:2024lkw}.
At the moment, 
the meaning of the prefactor  $1/(\eta q)$ in (\ref{1.8}) is unclear. Similar prefactors appear also in higher order giant graviton contributions 
and in more general Schur correlations functions \cite{Imamura:2024lkw,Beccaria:2024oif}.

\paragraph{New Results for large representations}

In this paper, we study the Schur 2-point function when the two Wilson lines are in the rank-$k$ symmetric $(k)$ or antisymmetric $[k]$ representation. 
Although the BPS Wilson line in fundamental representation may be represented  in terms of a fundamental string,
an alternative  description is by a D3 brane carrying electric flux and pinching off at the boundary of $AdS_{5}$ along the Wilson line 
\cite{Callan:1997kz,Gibbons:1997xz,Rey:1998ik,Drukker:2005kx}. 

This D-brane probe picture is  convenient when the fundamental representation is replaced by 
large higher  representations. In particular, a Wilson line in rank-$k$ symmetric representation corresponds to $k$ fundamental strings emerging as spikes on 
an extra D3-brane wrapping $AdS_{2}\times S^{2}\subset  AdS_{5}$, denoted $\text{D3}_{k}$  
\cite{Drukker:2005kx,Gomis:2006sb,Gomis:2006im,Rodriguez-Gomez:2006fmx,Yamaguchi:2007ps}. 
The case of rank-$k$ antisymmetric 
representation corresponds instead to $k$ fundamental strings attached to a D5-brane wrapping $AdS_{2}\times S^{4}\subset AdS_{5}\times S^{5}$, denoted $\text{D5}_{k}$
\cite{Yamaguchi:2006tq,Gomis:2006sb,Gomis:2006im,Rodriguez-Gomez:2006fmx,Hartnoll:2006is}. 
\footnote{More general representations and their D-brane probe description are discussed in [6].}

For the 2-point function in the symmetric or antisymmetric cases, we examine the leading giant graviton correction to 
line indices, \cf (\ref{1.4}),
\be
\I^{U(N)}_{S_{k}}(\eta; q) = \I^{U(N)}_{(k), \overline{(k)}}(\eta; q), \qquad
\I^{U(N)}_{A_{k}}(\eta; q) = \I^{U(N)}_{[k], \overline{[k]}}(\eta; q)\,.
\ee
We find that the ratios
\be
\la{1.11}
R_{\mathsf R}^{U(N)}(\eta; q) = \frac{\I^{U(N)}_{\mathsf R}(\eta; q)}{\I^{U(\infty)}_{\mathsf R}(\eta; q)}, \qquad \mathsf R = S_{k}, A_{k}\, ,
\ee
may be written at large $N$ and fixed $k$ as 
\bea
\la{1.12}
R_{S_{k}}^{U(N)}(\eta; q) &= 1+[\eta^{N}\,G_{S_{k}}^{+}(\eta; q)+\eta^{-N}\, G_{S_{k}}^{-}(\eta; q)]\, q^{N}+\cdots, \\
R_{A_{k}}^{U(N)}(\eta; q) &= 1+[\eta^{N-k}\,G_{A_{k}}^{+}(\eta; q)+\eta^{-(N-k)}\, G_{A_{k}}^{-}(\eta; q)]\, q^{N-k}+\cdots\, ,
\eea
where again $G^{-}_{\mathsf R}(\eta; q) = G^{+}_{\mathsf R}(\eta^{-1}; q)$ and, at small $q$, $G^{+}_{S_{k}}(\eta; q) = \mc O(1)$
and $G^{+}_{A_{k}}(\eta; q) = \mc O(q)$.
Dots in (\ref{1.12}) include in both cases higher giant graviton contributions of the form $q^{2N+\delta}$ where $\delta$ is a $k$-dependent integer. These
are beyond our scope and will not be discussed. Here, we present explicit formulas for the functions $G^{\pm}_{\mathsf R}(\eta; q)$.

From the finite $k$ results, we may examine the structure of the correction for large $k$.
In the symmetric case, when the functions $G^{\pm}_{S_{k}}(\eta; q)$ are expanded in powers of $q$,  $k$-dependent corrections appear  
taking the form of powers $q^{k-n}$ with positive $n$.
This implies that when $k\gg 1$, and irrespectively of its possible scaling with $N$, one has a definite $k\to\infty$ limit of the form 
\be
\la{1.13}
\lim_{k\to\infty}R_{S_{k}}^{U(N)}(\eta; q) \equiv R_{S}^{U(N)}(\eta; q) = 1+[\eta^{N}\,G_{S}^{+}(\eta; q)+\eta^{-N}\, G_{S}^{-}(\eta; q)]\, q^{N}+\cdots\, .
\ee
We prove that the asymptotic functions $G^{\pm}_{S}(\eta; q)$ are given by  the  exact residue formula
\ba
\la{1.14}
G^{+}_{S}(\eta; q) &= \frac{1}{\eta^{2} q}\ \mathop{\text{Res}}_{\eps=0}\bigg[\frac{1}{\eps^{3}}\frac{(\eps;\frac{q}{\eta})^{2}_{\infty}}{(\eps q^{2};\frac{q}{\eta})^{2}_{\infty}}
\frac{\vth(\eps\eta^{2}; \frac{q}{\eta})}{\vth(\eps; \frac{q}{\eta})}\bigg]\  G^{+}_{\rm D3}(\eta; q)\, ,
\ea
that evaluates to 
\ba
\la{ana}
G^{+}_{S}(\eta; q) &= \bigg[
\frac{1}{\eta q}+1+\frac{2}{\eta^{2}}+\frac{1}{\eta^{4}}+\bigg(\frac{2}{\eta^{5}}+\frac{1}{\eta^{3}}-\frac{2}{\eta}\bigg)\, q
+\bigg(\frac{5}{\eta^{6}}-\frac{2}{\eta^{4}}-\frac{3}{\eta^{2}}\bigg)\, q^{2}+\cdots
\bigg]\  G^{+}_{\rm D3}(\eta; q) \, .
\ea
In the antisymmetric case,  $G^{\pm}_{A_{k}}(\eta; q)$ has again powers of $q$ and $k$-dependent corrections as well. The latter take now the form  $q^{-k+n}$ with positive $n$,
and are  are dominant in the regime  $1\ll k\ll N$. They  determine the correction
\be
\la{antilim}
R_{S_{k}}^{U(N)}(\eta; q) \stackrel{1\ll k \ll N}{\equiv} R_{A}^{U(N)}(\eta; q) = 1+[\eta^{N-k}\,G_{A}^{+}(\eta; q)+\eta^{-(N-k)}\, G_{A}^{-}(\eta; q)]\, q^{N-k}+\cdots\, ,
\ee
where the functions $G_{A}^{\pm}(\eta; q)$ are $k$-independent. We prove that they are given exactly by 
\be
\la{1.16}
G_{A}^{+}(\eta; q) =  
-\frac{\eta q}{1-\eta q}\frac{(\frac{q}{\eta})_{\infty}}{(\frac{1}{\eta^{2}}; \frac{q}{\eta})_{\infty}}\,.
\ee
that we may write like in (\ref{ana}) as 
\be
G_{A}^{+}(\eta; q) =  \frac{1}{1-\eta q}\frac{(\eta q; \frac{q}{\eta})_{\infty}}{(\frac{q}{\eta})_{\infty}}\, G_{\rm D3}^{+}(\eta; q) = 
\bigg[
1+\frac{1}{\eta}\,q+\bigg(\frac{2}{\eta^{2}}-1\bigg)\, q^{2}+\bigg(\frac{3}{\eta^{3}}-\frac{2}{\eta}\bigg)\, q^{3}+\cdots
\bigg]\, G_{\rm D3}^{+}(\eta; q) \, .
\ee
Expressions (\ref{1.14}) and (\ref{1.16}) exhibit  explicit factorization of the undecorated index, similar to (\ref{1.8}), and are the main result of our analysis. 

They are expected to represent fluctuations of the probe $\text{D3}_{k}$ or $\text{D5}_{k}$
 branes in presence of the D3 giant graviton in the same spirit as \cite{Imamura:2024lkw} for the case of a pair of semi-infinite 
fundamental strings. 
In fact, for the antisymmetric $\text{D5}_{k}$ case, Imamura and Inoue have recently conjectured a quadruple-sum giant graviton expansion of the form  \cite{Imamura:2024pgp}
\be
\frac{\I_{A_{k}}^{U(N)}(\eta; q)}{I_{A_{\infty}}^{U(\infty)}(\eta; q)} = \sum_{m,m',n,n'\ge 0}(\eta q)^{km+(N-k)m'}(\eta^{-1}q)^{kn+(N-k)n'}\mc F_{m,n,m',n'}(\eta; q),
\ee
where some of the functions $\mc F$ are computed in analytic form and other are extracted from the $q$-series of the ratio in the l.h.s.
For the leading term in $1\ll k\ll N$ regime, their analysis implies the relation  
\ba
\la{1.18}
G_{A}^{+}(\eta; q) = \PE\bigg[\frac{\frac{1}{\eta q}-\eta^{-1}q}{1-\eta^{-1}q}\bigg],
\ea
where the plethystic exponential in the r.h.s. is  the analytic continuation of the half-index \cite{Dimofte:2011py} 
of $\N=4$ $U(1)$ Maxwell theory  living on a 3-dimensional disk which is a part of the giant graviton separated out by the probe $\text{D5}_{k}$ brane.
It is readily seen that (\ref{1.18}) agrees with (\ref{1.16}), that we obtained by an independent matrix model analysis of the Schur correlation function. This confirms the construction in \cite{Imamura:2024pgp}. 

It would be very interesting to compare our result (\ref{1.14}) for the symmetric representation at large $k$ with finite $N$ effects from 
fluctuations of the $\text{D3}_{k}$ probe in the presence of the giant graviton. This could be worth given the puzzling different structure of fluctuations 
even at infinite $N$ for the $\text{D3}_{k}$ and $\text{D5}_{k}$ probes \cite{Faraggi:2011bb,Faraggi:2011ge}, 
despite equality of the symmetric and antisymmetric Schur 2-point functions in this limit.

\paragraph{Plan of the paper} In Section \ref{sec:basic}, we present the explicit definitions of the Schur 2-point functions in 
symmetric an antisymmetric representation and discuss explicit finite $N,k$ data from their matrix model representation. In Sections \ref{sec:Ninf},
we discuss their $N\to	\infty$ limit and the form it takes when  $k$ is also large. Section \ref{sec:HS} uses the Hubbard-Stratonovich transformation to 
give a representation of the 2-point functions suitable for the analysis of their finite $N$ corrections. This is worked out in Section \ref{sec:sym}
in the case of the symmetric representation. In particular, in Section \ref{sec:large-k}, we discuss
the large $k$ form of the leading giant graviton correction (\ref{1.13}). Section \ref{sec:anti} treats the antisymmetric case.
The detailed derivations are presented in technical Appendices.

\section{Schur line correlators in $S_{k}$ or $A_{k}$ representation}
\la{sec:basic}

The matrix integral (\ref{1.4}) can be expressed as a contour integral over holonomies $\bz=(z_{1}, \dots, z_{N})$
\bea
\la{2.1}
\I^{U(N)}_{\mathsf R_{1}, \mathsf R_{2}, \dots}(\eta; q) & = {\oint}_{|\bz|=1}D^{N}\bz\, \prod_{n\ge 1}\chi_{\mathsf R_{n}}(\bz)\, 
\PE[ f(\eta; q)\chi_{\smb}(\bz)\chi_{\smb}(\bz^{-1})], \\
D^{N}\bz &= \frac{1}{N!}\prod_{n=1}^{N}\frac{dz_{n}}{2\pi i\, z_{n}}\prod_{n\neq m}\bigg(1-\frac{z_{n}}{z_{m}}\bigg)\, .
\eea
For the symmetric representation $(k)$ and antisymmetric $[k]$, the characters are 
\be
\la{2.2}
\chi_{(k)}(\bm{z}) = \sum_{1\le i_{1}\le i_{2}\le \cdots \le  i_{k}\le N}z_{i_{1}}z_{i_{2}}\cdots z_{i_{k}},\qquad
\chi_{[k]}(\bm{z}) = \sum_{1\le i_{1}<i_{2}<\cdots < i_{k}\le N}z_{i_{1}}z_{i_{2}}\cdots z_{i_{k}}\,.
\ee
In the following we will make extensive use of Young tableaux expansions. To set notation, we recall that 
a partition $\l$ of the positive integer $k$, denoted $\l\vdash k$, can be represented as a Young tableau with $\l_{n}$ blocks in its $n$-th row, \ie 
$\l=(\l_{1}, \l_{2}, \dots)$ with $\l_{1}\ge \l_{2} \ge \cdots$
or in frequency representation $1^{r_{1}}2^{r_{2}}\cdots$. 
The number of parts of the partition $\l$ is $\ell(\l) =\sum_{n}r_{n}$ (the number of non-zero $\l_{i}$) and is the number of rows
in the associated Young tableau. The weight of the partition $\l$ is $|\l| = \sum_{n}\l_{n} = \sum_{n}n r_{n}$, the number of blocks in the Young tableau,
and $\l\vdash k$ is same as $|\l|=k$. With this notation, we also define
\be
\z_{\l} = \prod_{n=1}^{\infty}r_{n}!\, n^{r_{n}}\,.
\ee
The characters (\ref{2.2}) may be given as sums over partitions $\l\vdash k$
\be
\la{2.4}
\chi_{(k)}(\bm{z}) = \sum_{\l\vdash k}\frac{1}{\z_{\l}}\chi_{\smb}(\bm{z})^{\l},\qquad
\chi_{[k]}(\bm{z}) = \sum_{\l\vdash k}\frac{(-1)^{k-\ell(\l)}}{\z_{\l}}\chi_{\smb}(\bm{z})^{\l}\,,
\ee
where we denoted, for a generic function $F(\bm z)$
\be
F(\bm z)^{\l} = \prod_{n\ge 1}F(\bm z^{\l_{n}}), \qquad \bm z^{m} = (z_{1}^{m}, \dots, z_{N}^{m}).
\ee
Given the set $\bm{A}=(A_{1}, A_{2}, \dots)$ ,  the relation between plethystic exponentiation and Young tableau expansion is 
\be
\PE[\bm{A}] = \exp\sum_{n\ge 1}\frac{1}{n}A_{n} = \sum_{\l}\frac{1}{\z_{\l}}\bm{A}^{\l}, \qquad \bm{A}^{\l} = \prod_{n\ge 1}A_{n}^{\l_{n}}.
\ee
It follows from this relation that 
\be
\la{2.8}
\PE[-\bm{A}] = \PE[\bm{A}]^{-1}=\sum_{\l}\frac{(-1)^{\ell(\l)}}{\z_{\l}}\bm{A}^{\l}\, .
\ee
Comparing with (\ref{2.4}) gives the well-known generating functions of symmetric and antisymmetric characters in the form 
\bea
S(x; \bm{z}) &= \sum_{k\ge 0}x^{k}\chi_{(k)}(\bm z) = \PE[x\, \chi_{\smb}(\bm z)] = \prod_{n=1}^{N}\frac{1}{1-x z_{n}}, \\
A(x; \bm{z}) &= \sum_{k\ge 0}(-1)^{k}\,x^{k}\chi_{[k]}(\bm z) = \PE[-x\, \chi_{\smb}(\bm z)] = \prod_{n=1}^{N}(1-x z_{n})\, .
\eea

\subsection{Finite $N$ matrix integral data}
\la{sec:data}

Explicit expressions for ratios (\ref{1.11}) with fixed $N$ and increasing $k$ are obtained by evaluation of (\ref{2.1}). For $U(2)$ gauge group we get 
\bea
\la{ex-s-2}
R_{S_{2}}^{U(2)}(\eta; q) &= 1-(1+\eta^{-2}+\eta^2) \,q^2-2\,(\eta^{-3}+\eta^{-1} +\eta + \eta^3) \,
q^3-(\eta^{-4}-5\eta^{-2}-8-5 \eta^2+\eta^4) \,q^4+\cdots, \\
R_{S_{3}}^{U(2)}(\eta; q) &= 1-(1+\eta^{-2}+\eta^2) \,q^2-2\,(\eta^{-3}+\eta^{-1} +\eta + \eta^3) \,
q^3-(2\eta^{-4}-3\eta^{-2}-5-3 \eta^2+2 \eta^4) \,q^4+\cdots, 
\eea
and same for $S_{k}$ with $k>3$ at this order in $q$.
For $U(3)$ gauge group
\bea
\la{ex-s-3}
R_{S_{2}}^{U(3)}&(\eta; q) = 1-(\eta^{-3}+\eta^{-1}+\eta +\eta ^3) \,
q^3-2(\eta^{-4}+\eta^{-2}+2+ \eta ^2+ \eta ^4)\, 
q^4\\
& -(\eta^{-5}-3\eta^{-3}-8\eta^{-1}-8 \eta -3 
\eta ^3+\eta ^5) \,q^5-(\eta^{-6}-3\eta^{-4}-5\eta^{-2}-5-5 \eta ^2-3 \eta ^4+\eta ^6)\, q^6+\cdots,  \\
R_{S_{3}}^{U(3)}&(\eta; q) = 1-(\eta^{-3}+\eta^{-1}+\eta +\eta ^3) \,
q^3-2(\eta^{-4}+\eta^{-2}+2+ \eta ^2+ \eta ^4)\, 
q^4\\
& -(2\eta^{-5}-\eta^{-3}-4\eta^{-1}-4 \eta -\eta 
^3+2 \eta ^5) \,q^5-(\eta^{-6}-4\eta^{-4}-8\eta^{-2}-13-8 \eta ^2-4 \eta ^4+\eta ^6) \,q^6+\cdots,  \\
R_{S_{4}}^{U(3)}&(\eta; q) = 1-(\eta^{-3}+\eta^{-1}+\eta +\eta ^3) \,
q^3-2(\eta^{-4}+\eta^{-2}+2+ \eta ^2+ \eta ^4)\, 
q^4\\
& -(2\eta^{-5}-\eta^{-3}-4\eta^{-1}-4 \eta -\eta 
^3+2 \eta ^5) \,q^5-2\,(\eta^{-6}-\eta^{-4}-2\eta^{-2}-4-2 \eta ^2- 
\eta ^4+ \eta ^6) \,q^6+\cdots,  
\eea
and same for $S_{k}$ with $k>4$ at this order in $q$. Continuing at higher $N$ shows that there is a well-defined limit
\be
\la{symlim}
R_{S}^{U(N)}(\eta; q) = \lim_{k\to\infty}R_{S_{k}}^{U(N)}(\eta; q).
\ee
The same analysis for the antisymmetric representation makes no sense since we cannot take $k$ large at fixed $N$. 
In this case the first correction is $\sim q^{N-k+1}$. 
For instance taking $N=5$ and $k=2,3,4,5$ one finds 
\ba
R_{A_{2}}^{U(5)}(\eta; q) &= 1-(\eta^{-4}+\eta^{-2}+1+\eta^{2}+\eta^{4})\, q^{4}
-(\eta^{-5}-\eta^{-1}-\eta-\eta^{5})\, q^{5}-(\eta^{-6}-1+\eta^{6})\,q^{6}+\cdots, \\
\la{2.17}
R_{A_{3}}^{U(5)}(\eta; q) &= 1-(\eta^{-3}+\eta^{-1}+\eta+\eta^{3})\,q^{3}-(\eta^{-4}+1+\eta^{4})\,q^{4}-(\eta^{-5}-\eta^{-1}-\eta+\eta^{5})\, q^{5} \\
& -(\eta^{-6}-\eta^{-4}-\eta^{-2}-1-\eta^{2}-\eta^{4}+\eta^{6})\, q^{6}+\cdots, \notag \\
R_{A_{4}}^{U(5)}(\eta; q) &= 1-(\eta^{-2}+1+\eta^{2})\, q^{2}-(\eta^{-3}+\eta^{3})\, q^{3}-(\eta^{-4}-\eta^{-2}-1-\eta^{2}+\eta^{4})\, q^{4}\\
& +(\eta^{-3}+\eta^{-1}+\eta+\eta^{3})\, q^{5}+(\eta^{-4}+\eta^{-2}+\eta^{2}+\eta^{4})\, q^{6}+\cdots, \notag \\
R_{A_{5}}^{U(5)}(\eta; q) &= 1-(\eta^{-1}+\eta )\, q-(\eta^{-2}-1+\eta^{2}) \,q^2
+(\eta^{-1}+\eta )\, q^3+q^4
+(\eta^{-5}+\eta^{5})\, q^5\lp
-(\eta^{-4}+\eta^{4})\, q^6+\cdots.
\ea
In the following, we will derive exact expressions for these expansions, up to terms of order $q^{2N+\cdots}$ where  higher order giant graviton effects
are to be taken into account. 

In particular, we will examine the dependence on the rank $k$.
At single giant graviton level, $k$-dependent corrections in symmetric case will contribute terms $q^{N+k+\cdots}$ (dots being fixed numerical integers)
and will thus be neglibile if $k$ is large, consistent with the limit  (\ref{1.13}). Instead, 
in antisymmetric case, $k$-dependent terms will be of the form $q^{N-k+\cdots}$
and will mix with $q^{N+\cdots}$ terms. They are disentangled in the limit $1\ll k \ll N$ and will be shown to be captured by (\ref{antilim}).

\subsection{ $N\to\infty$ limit at fixed $k$}
\la{sec:Ninf}

Let us begin with the  Schur  2-point function in the symmetric representation. According to (\ref{2.1}), we need
\ba
\la{2.20}
I_{S_{k}}^{U(N)}(\eta; q) = {\oint}_{|\bz|=1}D^{N}\bz\, \sum_{\l,\l'\vdash k}\frac{1}{\z_{\l}\z_{\l'}}\chi_{\smb}(\bm z)^{\l}\chi_{\smb}(\bm z^{-1})^{\l'}
\PE[ f(\eta; q)\chi_{\smb}(\bz)\chi_{\smb}(\bz^{-1})],
\ea
In the limit $N\to\infty$, only diagonal terms survive in the sum over $\l, \l'$ and we get 
\be
\la{2.21}
\I^{U(\infty)}_{S_{k}}(\eta; q) = \sum_{\l\vdash k}\frac{1}{\z_{\l}^{2}}\, \I_{\l}(\eta; q)\, ,
\ee
with 
\be
\la{2.22}
\I_{\l}(\eta; q) =\lim_{N\to\infty}
 {\oint}_{|\bz|=1}D^{N}\bz\, \chi_{\smb}(\bm z)^{\l}\chi_{\smb}(\bm z^{-1})^{\l}\,
\PE[ f(\eta; q)\chi_{\smb}(\bz)\chi_{\smb}(\bz^{-1})]\, .
\ee
Same is obtained starting from the antisymmetric representation because $[(-1)^{k-\ell(\l)}]^{2}=1$, \ie
\be
\I^{U(\infty)}_{S_{k}}(\eta; q) = \I^{U(\infty)}_{A_{k}}(\eta; q) \,.
\ee
To compute (\ref{2.22}) it is convenient to introduce a multi-coupling unitary matrix integral \cite{Murthy:2022ien} with $\bg = (g_{1}, g_{2}, \dots)$
\be
Z_{N}(\bg) = \int_{U(N)}dU\, \exp\bigg(\sum_{n=1}^{\infty}\frac{1}{n}g_{n}\Tr U^{n}\Tr U^{-n}\bigg) = 
 {\oint}_{|\bz|=1}D^{N}\bz\, \PE[ \bm{g}\,\chi_{\smb}(\bz)\chi_{\smb}(\bz^{-1})]\, .
\ee
It reduces to the Schur index by identification
\be
g_{n} = f(\eta^{n}; q^{n})\,.
\ee
For $N\to\infty$, one has   \cite{Murthy:2022ien} 
\be
\la{2.26}
Z_{\infty}(\bg) =  \prod_{n=1}^{\infty}\frac{1}{1-g_{n}}\,,
\ee
and thus the multi-coupling version of (\ref{2.22}) reads
\ba
\I_{\l}(\bg) =\l_{1} \frac{\partial}{\partial g_{\l_{1}}}\cdots\l_{\ell(\l)}\frac{\partial}{\partial g_{\l_{\ell(\l)}}}\prod_{n\ge 1}\frac{1}{1-g_{n}} = 
\prod_{m\ge 1}m^{r_{m}}\partial_{g_{m}}^{r_{m}}\prod_{n\ge 1}\frac{1}{1-g_{n}} = \prod_{n\ge 1}\frac{n^{r_{n}}r_{n}!}{(1-g_{n})^{r_{n}}}\times Z_{\infty}(\bg)\, .
\ea
Hence, the multi-coupling version of (\ref{2.21}), denoted $Z_{\infty}^{S_{k}}(\bg)$, obeys 
\ba
\la{2.28}
W_{k}(\bg) \equiv  \frac{Z_{\infty}^{S_{k}}(\bg) }{Z_{\infty}(\bg)} &= \sum_{\l\vdash k}\frac{1}{\z_{\l}^{2}}\, \prod_{n\ge 1}\frac{r_{n}!}{(1-g_{n})^{r_{n}}} 
= \sum_{\l\vdash k}\prod_{n\ge 1}\frac{1}{(r_{n}!)^{2}\,n^{2r_{n}}}\frac{n^{r_{n}}r_{n}!}{(1-g_{n})^{r_{n}}} \lp 
= \sum_{\l\vdash k}\prod_{n\ge 1}\frac{1}{r_{n}!\,n^{r_{n}}}\frac{1}{(1-g_{n})^{r_{n}}}  
= \exp\bigg[\sum_{n=1}^{\infty}\frac{1}{n}\frac{\eps^{n}}{1-g_{n}}\bigg]\bigg|_{\eps^{k}}.
\ea
This gives the following expression for the ratio at $N=\infty$ 
\ba
\la{2.29}
W_{k}(\eta; q) \equiv  \frac{\I_{S_{k}}^{U(\infty)}(\eta; q) }{\I^{U(\infty)}(\eta; q)} &= \exp\bigg[\sum_{n=1}^{\infty}\frac{1}{n}\frac{\eps^{n}}{1-f(\eta^{n}; q^{n})}\bigg]\bigg|_{\eps^{k}}
 = \PE\bigg[\frac{\eps}{1-f(\eta; q)}\bigg]\bigg|_{\eps^{k}}\,.
\ea
The result (\ref{2.29}) agrees with the conjectured formula Eq.~(5.13) in \cite{Hatsuda:2023iwi}, \ie
\be
\PE\bigg[\sum_{n=1}^{\infty}\frac{1}{n}\frac{\eps^{n}}{1-f(\eta^{n}; q^{n})}\bigg]\bigg|_{\eps^{k}} = \sum_{n=0}^{k}\frac{1}{(\eta q; \eta q)_{n}(\eta^{-1}q; \eta^{-1}q)_{k-n}}
- \sum_{n=0}^{k-1}\frac{1}{(\eta q; \eta q)_{n}(\eta^{-1}q; \eta^{-1}q)_{k-n-1}}\,.
\ee

\subsubsection{Taking $k\to \infty$}

When $g_{n}=f(q^{n})$ with $f(q) = \mc O(q)$ it is interesting to look at the large $k$ limit of (\ref{2.29}) within the  $q$-expansion. 
To this aim, we start from 
\ba
\PE\bigg[\frac{\eps}{1-f(\eta; q)}\bigg]\bigg|_{\eps^{k}} =\oint\frac{d\eps}{2\pi i\eps^{k+1} }\frac{1}{1-\eps}\PE\bigg[\eps\frac{f(\eta; q)}{1-f(\eta; q)}\bigg]\, .
\ea
If $k\to\infty$ at some order in $q$ expansion, the only pole that is relevant is $\eps=1$ because other poles give suppressed 
contributions $\sim q^{k+\cdots}$. Thus, \cf (\ref{2.29}),
\ba
\la{2.32}
\mc W(\eta; q) \equiv \lim_{k\to\infty} W_{k}(\eta; q) = \lim_{k\to\infty} \frac{\I_{S_{k}}^{U(\infty)}(\eta; q) }{\I^{U(\infty)}(\eta; q)} = \PE\bigg[\frac{f(\eta; q)}{1-f(\eta; q)}\bigg]\,,
\ea
and we remind that same is obtained using $A_{k}$. The expression (\ref{2.32}) agrees with the probe D3 brane computed in Eq.~(6.12) of \cite{Gang:2012yr}
since
\be
\frac{f(\eta;q)}{1-f(\eta;q)} = \frac{\eta q}{1-\eta q}+\frac{\eta^{-1}q}{1-\eta^{-1}q} =\sum_{n=1}^{\infty}(\eta^{n}+\eta^{-n})q^{n}\,.
\ee
This also gives
\ba
\lim&_{k\to\infty}\I^{U(\infty)}_{S_{k}}(\eta; q) = \PE\bigg[\frac{f(\eta; q)}{1-f(\eta; q)}\bigg]
\prod_{n=1}^{\infty}\frac{1}{1-f(\eta^{n};q^{n})} \lp
= \prod_{n\ge 1}\PE\bigg[(\eta^{n}+\eta^{-n})q^{n}-q^{2n}+(\eta^{n}+\eta^{-n})q^{n}] = 
\prod_{n=1}^{\infty}\frac{1-q^{2n}}{(1-\eta^{n} q^{n})^{2}(1-\eta^{-n}q^{n})^{2}}\, ,
\ea
in agreement with Eq.~(5.14) of \cite{Hatsuda:2023iwi}.

\subsubsection{Special limits}

In the following we will consider two special limit. The first is simply the unrefined case
\be
\la{2.35}
\text{unrefined limit:}\qquad \eta=1\,.
\ee
The second is the $\frac{1}{2}$-BPS limit
\be
\la{2.36}
\tfrac{1}{2}\text{-BPS limit:}\qquad \eta\, q=t, \qquad q\to 0\,.
\ee
For the key quantity $\mc W(\eta; q)$ we get 
\ba
\mc W(1; q) &= \PE\bigg[\frac{2q}{1-q}\bigg] = \prod_{n \ge 0}\PE[2q^{n+1}] =\prod_{n\ge 1}\frac{1}{(1-q^{n})^{2}}= \frac{1}{(q)_{\infty}^{2}},\\
\mc W(t) &\equiv \lim_{q\to 0}\mc W(t/q; q) = \PE\bigg[\frac{t}{1-t}\bigg] = \prod_{n\ge 1}\frac{1}{1-t^{n}} = \frac{1}{(t)_{\infty}}\,.
\ea

\subsection{Finite $N$ Schur line correlators from the HS functional}
\la{sec:HS}

At finite $N$, we need to keep non-diagonal terms in the double tableaux sum in (\ref{2.20}), \ie include contributions from  $\l'\neq \l$.
It is convenient to use the Hubbard-Stratonovich functional considered in \cite{Murthy:2022ien}
\be
\wt Z_{N}(\bm{t}^{+}, \bm{t}^{-}) = \int_{U(N)}dU\, \exp\bigg(\sum_{n=1}^{\infty}\frac{1}{n}(t_{n}^{+}\Tr U^{n}+t_{n}^{-} \Tr U^{-n})\bigg)\,.
\ee
For a function $f(\bm{t}^{+}, \bm{t}^{-})$, we define
\be
\la{2.40}
\langle f\rangle_{\bm{g}} = \int\prod_{n=1}^{\infty}\frac{dt^{+}_{n}dt^{-}_{n}}{2\pi\, n g_{n}}e^{-\frac{1}{ng_{n}}t^{+}_{n}t^{-}_{n}}\, f(\bm{t}^{+}, \bm{t}^{-})\,,
\qquad 
\int\frac{dt^{+}dt^{-}}{2\pi g}e^{-\frac{1}{g}t^{+}t^{-}}(t^{+})^{n_{+}}(t^{-})^{n_{-}} = n!\, g^{n}\, \delta_{n^{+}, n^{-}}\,.
\ee
Differentiating, we get 
\ba
\la{2.41}
Z_{N}^{S_{k}}(\bg) =  \sum_{\l,\l'\vdash k}\frac{1}{\z_{\l}\z_{\l'}} \int\prod_{n\ge 1}\frac{dt^{+}_{n}dt^{-}_{n}}{2\pi\, n g_{n}}e^{-\frac{1}{ng_{n}}t^{+}_{n}t^{-}_{n}}
n^{r_{n}+r'_{n}}\partial_{t_{n}^{+}}^{r_{n}}\partial_{t_{n}^{-}}^{r'_{n}}\, \wt Z_{N}(\bm{t}^{+}, \bm{t}^{-})\, .
\ea
The previous result (\ref{2.28}) for $N\to\infty$ is easily recovered as a special case. To see this, we use 
\be
\wt Z_{\infty}(\bm{t}^{+}, \bm{t}^{-}) = \exp\bigg(\sum_{n=1}^{\infty}\frac{1}{n}t_{n}^{+}t_{n}^{-}\bigg)\,,
\ee
and relation (\ref{B.7}) with $X=1/n$ gives
\ba
Z_{\infty}^{S_{k}}(\bg) &=  \sum_{\l,\l'\vdash k}\frac{1}{\z_{\l}\z_{\l'}} \int\prod_{n\ge 1}\frac{dt^{+}_{n}dt^{-}_{n}}{2\pi\, n g_{n}}e^{-\frac{1}{ng_{n}}t^{+}_{n}t^{-}_{n}}
n^{r_{n}+r'_{n}}
e^{\frac{1}{n}t^{+}_{n}t^{-}_{n}}(t^{+}_{n})^{r'_{n}}(t^{-}_{n})^{r_{n}}n^{-r_{n}-r'_{n}}\lp
 \sum_{p\ge 0}p!\binom{r_{n}}{p}\binom{r'_{n}}{p} \bigg(\frac{1}{n}t^{+}_{n}t^{-}_{n}\bigg)^{-p}\, .
\ea
Integrating over $\bm{t}^{\pm}$, we get a nonzero result if $\l=\l'$ so 
\ba
Z_{\infty}^{S_{k}}(\bg) &=  \sum_{\l\vdash k}\frac{1}{\z_{\l}^{2}} \int\prod_{n\ge 1}\frac{dt^{+}_{n}dt^{-}_{n}}{2\pi\, n g_{n}}e^{-\frac{1-g_{n}}{ng_{n}}
t^{+}_{n}t^{-}_{n}}
 \sum_{p\ge 0}n^{p}p!\binom{r_{n}}{p}\binom{r_{n}}{p} (t^{+}_{n}t^{-}_{n})^{r_{n}-p}\lp
 =  Z_{\infty}(\bg)\,\sum_{\l\vdash k}\frac{1}{\z_{\l}^{2}}n^{r_{n}}
 \sum_{p\ge 0}p!\binom{r_{n}}{p}\binom{r_{n}}{p}(r_{n}-p)! \bigg(\frac{g_{n}}{1-g_{n}}\bigg)^{r_{n}-p} \lp
 = Z_{\infty}(\bg)\,\sum_{\l\vdash k}\frac{1}{\z_{\l}^{2}}n^{r_{n}}r_{n}!(1-g_{n})^{-r_{n}} = 
 Z_{\infty}(\bg) \sum_{\l\vdash k}\prod_{n\ge 1}\frac{1}{n^{r_{n}}r_{n}!}\frac{1}{(1-g_{n})^{r_{n}}}\, ,
\ea
where we applied (\ref{2.40}) with coupling $g = ng_{n}/(1-g_{n})$. This is same as (\ref{2.28}).

\section{Single giant graviton correction: Symmetric representation}
\la{sec:sym}

The leading giant graviton correction to (\ref{2.41}) follows from the correction to $\wt Z_{N}(\bm{t}^{+}, \bm{t}^{-})$
computed in \cite{Murthy:2022ien}. The main result is a determinantal expansion with first term 
\ba
\frac{\wt Z_{N}(\bm{t}^{+}, \bm{t}^{-})}{\wt Z_{\infty}(\bm{t}^{+}, \bm{t}^{-})} = 1-K_{N}(\bm{t}^{+}, \bm{t}^{-})+\cdots, \qquad
K_{N}(\bm{t}^{+}, \bm{t}^{-}) = \mathop{\sum_{N<r}}_{r\in\mathbb{Z}+\frac{1}{2}} \widetilde{K}(r,r; \bm{t}^{+}, \bm{t}^{-})\,,
\ea
where 
\bea
\la{3.2}
& \sum_{r,s\in\mathbb{Z}+\frac{1}{2}}\widetilde K(r,s; \bm{t}^{+}, \bm{t}^{-})\, z^{r}w^{-s} = \frac{J(z; \bm{t}^{+}, \bm{t}^{-})}{J(w; \bm{t}^{+}, \bm{t}^{-})}\, \frac{\sqrt{zw}}{z-w}, 
\qquad |w|<|z|,\\
& J(z; \bm{t}^{+}, \bm{t}^{-}) = \exp\bigg(\sum_{n=1}^{\infty}\frac{1}{n}(t_{n}^{+}z^{n}-t_{n}^{-}z^{-n})\bigg)\,. 
\eea
Thus, the correction to (\ref{2.41}) reads
\bea
\la{3.3}
Z_{N}^{S_{k}}(\bg) &=  Z_{\infty}^{S_{k}}(\bg) - \sum_{\l,\l'\vdash k}\frac{1}{\z_{\l}\z_{\l'}}\delta Z_{N}^{\l,\l'}(\bg)+\cdots,\\
\delta Z_{N}^{\l,\l'}(\bg) &=   \int\prod_{n\ge 1}\frac{dt^{+}_{n}dt^{-}_{n}}{2\pi\, n g_{n}}e^{-\frac{1}{ng_{n}}t^{+}_{n}t^{-}_{n}}
n^{r_{n}+r'_{n}}\partial_{t_{n}^{+}}^{r_{n}}\partial_{t_{n}^{-}}^{r'_{n}}\, 
[\wt Z_{\infty}(\bm{t}^{+}, \bm{t}^{-})K_{N}(\bm{t}^{+}, \bm{t}^{-})]+\cdots\, .
\eea
As we fully discuss in Appendix \ref{app:full}, one can prove the following result for the leading correction to the 
ratio $R_{S_{k}}^{U(N)}(\bg)$
\ba
\la{3.4}
R_{S_{k}}^{U(N)}(\bg) &= 1+\frac{1}{W_{k}(\bg)}\, \Phi^{S}_{k}(\bg; \zeta) G(\bg; \zeta)\bigg|_{\zeta^{-N}}+\cdots\, ,
\ea
with 
\ba
\la{3.5}
G(\bg; \zeta) &=  \frac{-\zeta}{(1-\zeta)^{2}} \PE\bigg[-\frac{\bg}{1-\bg}(1-\zeta)(1-\zeta^{-1})\bigg], \\
\la{3.6}
 \Phi_{k}^{S}(\bg; \zeta) &= \bigg[
 \frac{1}{1-\eps_{+}\eps_{-}}\frac{1-\eps_{+}}{1-\eps_{-}}\frac{1-\zeta\eps_{-}}{1-\zeta^{-1}\eps_{+}}\ \PE\bigg[
\frac{\bg}{1-\bg}\bigg(\eps_{+}\eps_{-}+\eps_{-}(1-\zeta)-\eps_{+}(1-\zeta^{-1})\bigg)\bigg]
 \bigg]_{\eps_{+}^{k},\eps_{-}^{k}} \,.
\ea
At fixed $k$, we get a sum of terms where the dependence on $\zeta$ is in simple powers. Each power $\zeta^{p}$ in 
\be
\frac{1}{W_{k}(\bg)}\, \Phi_{k}^{S}(\bg; \zeta) \, ,
\ee
simply replaces $G_{N}(\bg)\to G_{N-p}(\bg)$ in (\ref{3.4}), where $G_{N}(\bg) = G(\bg; \zeta)|_{\zeta^{-N}}$.

 \subsection{Explicit ratios $R_{S_{k}}^{U(N)}$ at fixed $k$ and special limits}
 \la{sec:lowk}

Evaluation of (\ref{3.4}) at fixed low $k$ is straightforward and reproduces the associated terms in the expansions
of $R_{S_{k}}^{U(N)}$ presented in Section \ref{sec:data}. 
To give explicit examples, for the rank-2, 3, 4 representations one finds, \cf (\ref{1.12}),
\ba
G^{+}_{S_{2}}(\eta;q) &= \frac{\eta^{2}}{1-\eta^{2}}+\frac{2(1+\eta^{4})}{\eta(1-\eta^{2})}\,q \notag\\
&+\frac{2-3\eta^{4}-4\eta^{6}+\eta^{8}}{\eta^{4}\,(1-\eta^{2})}\,q^{2}
+\frac{2+\eta^{2}-3\eta^{4}-3\eta^{6}-\eta^{8}-2\eta^{10}+\eta^{12}}{\eta^{7}(1-\eta^{2})}\,q^{3}+\cdots, \notag \\
G^{+}_{S_{3}}(\eta;q) &= \frac{\eta^{2}}{1-\eta^{2}}+\frac{2(1+\eta^{4})}{\eta(1-\eta^{2})}\,q  \\
&+\frac{2+\eta^{2}-\eta^{4}-3\eta^{6}+2\eta^{8}}{\eta^{4}\,(1-\eta^{2})}\,q^{2}
+\frac{2+2\eta^{2}-\eta^{4}-6\eta^{6}-2\eta^{8}-3\eta^{10}+\eta^{12}}{\eta^{7}(1-\eta^{2})}\,q^{3}+\cdots, \notag \\
G^{+}_{S_{4}}(\eta;q) &= \frac{\eta^{2}}{1-\eta^{2}}+\frac{2(1+\eta^{4})}{\eta(1-\eta^{2})}\,q+\frac{2+\eta^{2}-\eta^{4}-3\eta^{6}+2\eta^{8}}{\eta^{4}\,(1-\eta^{2})}\,q^{2}
+\frac{2(1+\eta^{2}-2\eta^{6}-\eta^{10}+\eta^{12})}{\eta^{7}(1-\eta^{2})}\,q^{3}+\cdots\,, \notag
\ea
where the above series are from exact expression obtained by (\ref{3.4}-\ref{3.6}) and we just wrote the first terms. Plugging the above into (\ref{1.12})
reproduces the expansions (\ref{ex-s-2}), (\ref{ex-s-3}). In particular, the first two terms in $G^{+}_{S_{k}}$ do not depend on $k$ and give
\be
R^{U(N)}_{S_{k}} = 1-\bigg[\frac{\eta(\eta^{-N-1}-\eta^{N+1})}{1-\eta^{2}}+\frac{2(1+\eta^{4})(\eta^{-N}-\eta^{-N})}{\eta(1-\eta^{2})}\,q+\cdots\bigg]\,q^{N}+\cdots\, .
\ee

It is also possible to consider the generalization of the subtracted ratio in (\ref{1.7}) corresponding to $k=1$, \ie we can 
introduce 
\be
\frac{\I_{S_{k}}^{U(N)}-W_{k}\,\I^{U(N)}}{\I^{U(\infty)}}  = 1+\bigg(\mathsf G^{+}_{S_{k}}(\eta; q)\, \eta^{N}+\mathsf G^{-}_{S_{k}}(\eta; q)\,\eta^{-N}\bigg)\, q^{N}+\cdots.
\ee
From the leading correction to the undecorated Schur index,
\bea
\la{3.9}
\frac{\I^{U(N)}(\eta; q)}{\I^{U(\infty)}(\eta; q)}&= 
1 +\bigg[\eta^{N}G^{+}_{\rm D3}(\eta; q)+\eta^{-N}G^{-}_{\rm D3}(\eta; q)\bigg]\, q^{N}+\mc O(q^{2N})\, , \\
G^{+}_{\rm D3}(\eta; q) &= G^{-}_{\rm D3}(\eta^{-1}; q) = -\eta^{2}q\, \frac{(\frac{q}{\eta})_{\infty}^{3}}{\vth(\eta^{2},\frac{q}{\eta})} = 
\PE\bigg[\frac{\frac{1}{\eta q}-\frac{2}{\eta}q+q^{2}}{1-\frac{q}{\eta}}\bigg]\, ,
\eea
we can write 
\ba
\mathsf G^{+}_{S_{k}}  &=X^{+}_{S_{k}}G^{+}_{\rm D3},\qquad 
X_{S_{k}}^{+} = W_{k}\bigg(\frac{G_{S_{k}}^{+}}{G^{+}_{\rm D3}}-1\bigg).
\ea
In analogy with the $k=1$ case, the quantity $X_{S_{k}}^{+}$ represents the contributions from fluctuations of the probe D3 brane
that is used to represent the $k$ coinciding fundamental strings building $S_{k}$, and attached to the giant graviton. 
For $k=1$, \cf (\ref{1.8}) and (\ref{1.9}), it was  \cite{Beccaria:2024oif} 
\be
X_{S_{1}}^{+} = \frac{1}{\eta q}\frac{(1-q^{2})^{2}}{(1-\eta^{-1}q)^{2}} = \frac{1}{\eta q}\, \PE[f_{\rm F}(\eta; q)]\, .
\ee
For $k>1$ exact expressions are unwieldy and little instructive. For instance, for $k=2$ we get 
\ba
X_{S_{2}}^{+} &= \frac{\eta^{3}+(1+\eta^{2})\,q-\eta(4+\eta^{2})\, q^{2}-(1-2\eta^{2})\,q^{3}+\eta(1-\eta^{2})\,q^{4}+q^{5}}{\eta^{4}q}\, \lp
\qquad \times \frac{(1-q^{2})^{2}}
{(1-\eta q)(1-\eta^{-1}q)^{2}(1-\eta^{-2}q^{2})^{2}}\,.
\ea
The second line may be written as a  simple plethystic exponential, but we also have a non trivial sum of monomials in the first line. 
A  similar result was found in \cite{Beccaria:2024oif}
for the symmetric 2-power of the fundamental representation.

The result in  $\frac{1}{2}$-BPS limit (\ref{2.36}) is much simpler and  one finds the following pattern,  valid for all $k$,
\be
\la{3.13}
X_{S_{k}}^{+} = \frac{1}{t}\prod_{p=1}^{k-1}\frac{1}{1-t^{p}} = \frac{1}{t\, (t; t)_{k-1}}\,.
\ee

 \subsection{Large $k$ limit and $R_{S}^{U(N)}$}
\la{sec:large-k}
  
 Taking into account that $g_{n} \sim q^{n}$, it will be convenient to organize expressions in powers of the auxiliary counting variable 
 $\kappa$ with $g_{n}\to \kappa^{n}g_{n}$. 
 The explicit form of $\Phi_{k}^{S}(\bg; \zeta)$ turns out to be  
\ba
\la{3.14}
\Phi_{k}^{S}(\bg; \zeta) &= \zeta -\zeta ^{1-k}+\zeta ^{-k}\lp
+(\zeta  -3 \zeta ^{1-k} +\zeta ^{2-k} +2 \zeta ^{-k} )\,g_{1}\, \kappa \lp
+[g_1^2+\zeta ^2 
g_1^2+\zeta ^{1-k} (-7 g_1^2-g_2)+\frac{1}{2} \zeta ^{2-k} (7 
g_1^2-g_2)+\frac{1}{2} \zeta  (-g_1^2+g_2)+\frac{1}{2} \zeta ^{3-k} 
(-g_1^2+g_2)\lp
+\zeta ^{-k} (4 g_1^2+g_2)]\, \kappa^{2}+\cdots\,.
\ea
Alternatively, after specializing $g_{n}=f(\eta^{n}; q^{n})$, we can  organize the expansion in powers of $q$. Defining 
\be
\Phi_{k}^{S}(\eta; q; \zeta) = \sum_{p=0}^{\infty}\Phi_{k,p}^{S}(\eta; \zeta)\, q^{p},
\ee
we have 
\bea
\la{3.16}
\Phi_{k,0}^{S}(\eta; \zeta) &= \zeta^{-k}(1-\zeta+\zeta^{1+k}), \\
\Phi_{k,1}^{S}(\eta; \zeta) &= (\eta+\eta^{-1})\zeta^{-k}(2-3\zeta+\zeta^{2}+\zeta^{k+1}), \\
\Phi_{k,2}^{S}(\eta; \zeta) &= (\eta^{-2}+2+\eta^{2})(1+\zeta^{2})-3 \zeta -\zeta ^{3-k}+\zeta ^{2-k} (3\eta^{-2}+5+3 
\eta ^2)\\
& +\zeta ^{-k} (5\eta^{-2}+4+5 \eta ^2)-8 \zeta^{1-k} (\eta^{-2}+1+\eta^{2}), \\
& \cdots\, .
\eea
The above coefficients are valid for large enough $k$. In more details, we need $k\ge p$ in $\Phi_{k,p}^{S}(\eta; \zeta)$. 
We can now take the $k\to\infty$ limit at fixed $N$. We can drop
from $\Phi_{k,p}^{S}$ powers of $\zeta$ with exponent dependent on $k$ since these are of the form  $q^{k+\cdots}$ and give terms suppressed at large $k$.
Special large $k$ scaling relations like  $k\sim N$ or $k\sim N^{2}$ are equivalent from this point of view.

To isolate powers of $\zeta$ with exponent independent on $k$, we start from  
\ba
 \Phi_{k}^{S}(\bg; \zeta) &= \oint\frac{d\eps_{+}}{2\pi i \eps_{+}^{k+1}}\oint\frac{d\eps_{-}}{2\pi i \eps_{-}^{k+1}}
 \frac{1}{1-\eps_{+}\eps_{-}}\frac{1-\eps_{+}}{1-\eps_{-}}\frac{1-\zeta\eps_{-}}{1-\zeta^{-1}\eps_{+}}\ \lp
\qquad \qquad  \PE\bigg[
\frac{\bg}{1-\bg}\bigg(\eps_{+}\eps_{-}+\eps_{-}(1-\zeta)-\eps_{+}(1-\zeta^{-1})\bigg)\bigg]\,.
 \ea
 We integrate over $\eps_{-}$ by  taking minus the residues at $\eps_{-}=1,\eps_{+}^{-1}$. The first pole leaves a $k$ dependence. So, taking the residue 
 at the second pole, we define
\ba
\wt\Phi^{S}(\bg; \zeta) &= \zeta\,\oint\frac{d\eps}{2\pi i \eps}
\ \PE\bigg[\frac{\bg}{1-\bg}\bigg(1+\eps^{-1}(1-\zeta)-\eps(1-\zeta^{-1})\bigg)\bigg]\,.
 \ea
%
One checks that indeed it is correct, \ie generates all terms in (\ref{3.14}) that are $k$-independent
\ba
\wt\Phi^{S}(\bg; \zeta) &= \zeta +\zeta\,g_{1}\, 
+[g_1^2+\zeta ^2 
g_1^2+\frac{1}{2} \zeta  (-g_1^2+g_2)]+\cdots\,.
\ea
Each power $\zeta^{p}$ simply shifts $G_{N}(\eta; q)\to G_{N-p}(\eta; q)$ where, \cf (\ref{3.9}), 
\be
G_{N}(\eta; q) = -\bigg[\eta^{N+2}\frac{(\frac{q}{\eta})_{\infty}^{3}}{\vth(\eta^{2},\frac{q}{\eta})}
+\eta^{-N-2}\frac{(\eta\,q)_{\infty}^{3}}{\vth(\eta^{-2},\eta\,q)}\bigg]\,q^{N+1}.
\ee
This gives
\be
R_{S}^{U(N)}(\eta; q) = 1+[\eta^{N}G_{S}^{+}(\eta; q)+\eta^{-N}G_{S}^{-}(\eta; q)]\,q^{N}+\cdots, \qquad G_{S}^{-}(\eta; q) = G_{S}^{+}(\eta^{-1}; q), 
\ee
with
\ba
\frac{G_{S}^{+}(\eta; q)}{G_{\rm D3}^{+}(\eta; q)} &= \PE\bigg[-\frac{f}{1-f}\bigg]\frac{1}{\eta q}\,\oint\frac{d\eps}{2\pi i \eps}
\ \PE\bigg[\frac{f}{1-f}\bigg(1+\eps^{-1}(1-\eta^{-1} q^{-1})-\eps(1-\eta q)\bigg)\bigg]\, \lp
= \frac{1}{\eta q}\,\oint\frac{d\eps}{2\pi i \eps}
\ \PE\bigg[\frac{f}{1-f}\bigg(\frac{1}{\eps}\bigg(1-\frac{1}{\eta q}\bigg)-\eps(1-\eta q)\bigg)\bigg]\,. 
\ea
The plethystic exponential may be expressed in terms of $q$-Pochhammer symbols
\ba
\la{3.23}
&\PE\bigg[\frac{f}{1-f}\bigg(\frac{1}{\eps}\bigg(1-\frac{1}{\eta q}\bigg)-\eps(1-\eta q)\bigg)\bigg] = \PE\bigg[-
\frac{(1+\eta q \eps^{2})(1-2\eta q+\eta^{2})}{\eps\eta^{2}(1-\eta^{-1}q)}
\bigg] \lp
=\PE[\sum_{n=0}^{\infty}(-\frac{1}{\eps}+2q^{2}\eps-\frac{1}{\eps\eta^{2}}+\frac{2q}{\eps \eta}-\frac{q\eps}{\eta}-q\eta \eps)(\eta^{-1}q)^{n}]\lp
= \frac{(\eps^{-1}; \eta^{-1}q)_{\infty}\,(\eps^{-1}\eta^{-2}; \eta^{-1}q)_{\infty}\,
(\eps\eta^{-1}q; \eta^{-1}q)_{\infty}\,(\eps \eta q; \eta^{-1}q)_{\infty}}{(\eps q^{2}; \eta^{-1}q)^{2}_{\infty}\,
(\eps^{-1}\eta^{-1}q; \eta^{-1}q)_{\infty}^{2}}
\ea
The general ratio is thus
\be
\frac{G^{+}_{S}(\eta; q)}{G^{+}_{\rm D3}(\eta; q)} = \frac{1}{\eta q}\oint
\frac{d\eps}{2\pi i \eps}\frac{(\eps^{-1}; \eta^{-1}q)_{\infty}\,(\eps^{-1}\eta^{-2}; \eta^{-1}q)_{\infty}\,(\eps\eta^{-1}q; \eta^{-1}q)_{\infty}\,
(\eps \eta q; \eta^{-1}q)_{\infty}}{(\eps q^{2}; \eta^{-1}q)^{2}_{\infty}\,(\eps^{-1}\eta^{-1}q; \eta^{-1}q)^{2}_{\infty}}\,.
\ee
This may be written in a more compact form by using  $q$-theta functions, \cf (\ref{A.3}),
\be
\la{cont}
\frac{G^{+}_{S}(\eta; q)}{G^{+}_{\rm D3}(\eta; q)} = \frac{1}{\eta q^{2}}\oint
\frac{d\eps}{2\pi i \eps}\frac{(\eps;\frac{q}{\eta})^{2}_{\infty}}{(\eps q^{2};\frac{q}{\eta})^{2}_{\infty}}
\frac{\vth(\eps; \frac{q}{\eta})\, \vth(\eps\eta^{2}; \frac{q}{\eta})}{\vth(\eps\frac{\eta}{q}; \frac{q}{\eta})^{2}}.
\ee
From (\ref{A.4}), we have 
\be
\vth\bigg(\eps\frac{\eta}{q}; \frac{q}{\eta}\bigg) = -(q/\eta)^{-1/2}\eps\,\vth\bigg(\eps;\frac{q}{\eta}\bigg),
\ee
and thus we get \footnote{It would be interesting to see if methods of \cite{Pan:2021mrw} may deal with the contour integral (\ref{cont}) to give an explicit $\eta$ dependent
$q$-series, although this is not essential for our applications.}
\be
\frac{G^{+}_{S}(\eta; q)}{G^{+}_{\rm D3}(\eta; q)} = \frac{1}{\eta^{2} q}\oint
\frac{d\eps}{2\pi i \eps}\frac{1}{\eps^{2}}\frac{(\eps;\frac{q}{\eta})^{2}_{\infty}}{(\eps q^{2};\frac{q}{\eta})^{2}_{\infty}}
\frac{\vth(\eps\eta^{2}; \frac{q}{\eta})}{\vth(\eps; \frac{q}{\eta})}.
\ee
The explicit expansion in powers of $q$ is straightforward and reads
\ba
G & _{S}^{+}(\eta; q) = \frac{\eta ^2}{1-\eta ^2}+\frac{2 (1+\eta ^4) }{\eta (1-\eta^2)}\,q
+\frac{2+\eta ^2-\eta ^4-3 \eta ^6+2 \eta ^8}{\eta ^4(1-\eta ^2)}\,q^{2}\lp
+(2\eta^{-7}+4 \eta ^{-5}+4 \eta ^{-3}-2 \eta ^{3})\, q^3\lp
+(2\eta^{-10}+4\eta^{-8}+7\eta^{-6}-3\eta^{-4}-3\eta^{-2}+1-2 \eta ^4)\,q^{4}\lp
+(2\eta^{-13}+4\eta^{-11}+8\eta^{-9}-6\eta^{-5}+2\eta^{-3}-2\eta^{5})\, q^{5}+\cdots.
\ea

\subsubsection{Unflavored limit}

In unflavored limit $\eta\to 1$, the plethystic exponential (\ref{3.23}) simplifies to 
\ba
& \PE\bigg[-
\frac{(1+ q \eps^{2})(1-2 q+1)}{\eps(1-q)}
\bigg] = \PE\bigg[-\frac{2}{\eps}-2q\eps\bigg] = \bigg(1-\frac{1}{\eps}\bigg)^{2}(1-q\eps)^{2},
\ea
and thus
\ba
\lim_{\eta\to 1}\frac{G^{+}_{S}(\eta; q)}{G^{+}_{\rm D3}(\eta; q)} = \frac{1}{q}\oint\frac{d\eps}{2\pi i \eps}\bigg(1-\frac{1}{\eps}\bigg)^{2}(1-q\eps)^{2}
= q^{-1}+4+q.
\ea
The  first correction in $\eta-1$ can be computed as follows
\ba
\frac{1}{\eta q}& \,
\ \PE\bigg[\frac{f}{1-f}\bigg(\frac{1}{\eps}\bigg(1-\frac{1}{\eta q}\bigg)-\eps(1-\eta q)\bigg)\bigg]\lp
=\frac{1}{q}\bigg(1-(\eta-1)+\cdots\bigg) \PE\bigg[-\frac{2}{\eps}-2q\eps+\frac{2(1+\eps^{2}q^{2})}{\eps(1-q)}(\eta-1)+\cdots\bigg] \lp
= \frac{1}{q} \bigg(1-\frac{1}{\eps}\bigg)^{2}(1-q\eps)^{2}\bigg[1+\bigg(-1+\sum_{n=0}^{\infty}\bigg(\frac{2q^{n}}{\eps-q^{n}}+\frac{2\eps q^{n+2}}{1-\eps q^{n+2}}\bigg)\bigg)(\eta-1)+\cdots\bigg]\lp
=  \frac{1}{q} \bigg(1-\frac{1}{\eps}\bigg)^{2}(1-q\eps)^{2}\bigg[1+\bigg(-1+2\sum_{n,p=0}^{\infty}\bigg(\frac{q^{n}}{\eps}\frac{q^{np}}{\eps^{p}}+\eps q^{n+2}(\eps q^{n+2})^{p}
\bigg)\bigg)(\eta-1)+\cdots\bigg]\lp
=  \frac{1}{q} \bigg(1-\frac{1}{\eps}\bigg)^{2}(1-q\eps)^{2}\bigg[1+\bigg(-1+\sum_{p=0}^{\infty}
\frac{2(\eps^{-1-p}+\eps^{1+p}q^{2+2p})}{1-q^{p+1}}
\bigg)(\eta-1)+\cdots\bigg].
\ea
The coefficient of $\eps^{0}$ receives contributions from a finite number of terms in the sum over $p$, and this gives
\be
\frac{G^{+}_{S}(\eta; q)}{G^{+}_{\rm D3}(\eta; q)} = q^{-1}+4+q-\frac{1+8q+10q^{2}+8q^{3}+q^{4}}{q(1-q^{2})}(\eta-1)+\mc O((\eta-1)^{2})\,.
\ee
Using the near unflavored expansion \cite{Beccaria:2024szi}
\be
G^{+}_{\rm D3}(\eta; q) =-\frac{1}{2}q\,\frac{1}{\eta-1}-\frac{5}{4}q+\mc O((\eta-1)^{2}), 
\ee
gives the total correction
\be
\eta^{N}G_{S}^{+}(\eta; q)+\eta^{-N}G_{S}^{-}(\eta; q) \stackrel{\eta\to 1}{=} -N(1+4q+q^{2})-\frac{1-10q^{2}-16q^{3}-3q^{4}}{1-q^{2}}.
\ee
Extra powers of $N$ in the unrefined limit is a wall-crossing effect 
 \cite{Gaiotto:2021xce,Lee:2022vig,Beccaria:2023zjw}. They are expected to come on gravity side from zero mode of fluctuations
 \cite{Beccaria:2023cuo,Beccaria:2024vfx,Gautason:2023igo}.

\subsubsection{$\frac{1}{2}$-BPS limit}

In $\frac{1}{2}$-BPS limit (\ref{2.36}),  the plethystic exponential (\ref{3.23}) 
is
\ba
\PE\bigg[-
\frac{(1+\eta q \eps^{2})(1-2\eta q+\eta^{2})}{\eps\eta^{2}(1-\eta^{-1}q)}\bigg] = \PE\bigg[-\frac{1}{\eps}-t\eps\bigg] = \bigg(1-\frac{1}{\eps}\bigg)(1-t\eps),
\ea
and thus
\ba
\frac{G^{+}_{S}(t)}{G^{+}_{\rm D3}(t)} = \frac{1}{t}\oint\frac{d\eps}{2\pi i \eps} \bigg(1-\frac{1}{\eps}\bigg)(1-t\eps)
= \frac{1+t}{t}.
\ea
Using
\be
\la{3.37}
G^{+}_{\rm D3}(t) = -\frac{t}{1-t},
\ee
gives
\be
G^{+}_{S}(t) = -\frac{1+t}{1-t}.
\ee
This is consistent with (\ref{3.13}). Indeed, for large $k$
\be
X^{+}_{S} = W_{\infty}\bigg(\frac{G^{+}_{S}}{G^{+}_{\rm D3}}-1\bigg) = \bigg(\frac{1+t}{t}-1\bigg)\prod_{n\ge 1}\frac{1}{1-t^{n}} 
= \frac{1}{t}\prod_{n\ge 1}\frac{1}{1-t^{n}} = \lim_{k\to\infty}X^{+}_{S_{k}}.
\ee

\section{Single giant graviton correction: Antisymmetric representation}
\la{sec:anti}

For the antisymmetric representation, \cf (\ref{2.4}), we can repeat  the derivation in Appendix \ref{app:fullplet}
by taking into account relation (\ref{2.8}). This readily gives, \cf (\ref{3.4}),
\ba
R_{A_{k}}^{U(N)}(\bg) &= 1+\frac{1}{W_{k}(\bg)}\, \Phi_{k}^{A}(\bg; \zeta) G(\bg; \zeta)\bigg|_{\zeta^{-N}}+\cdots\, ,
\ea
with 
\ba
 \Phi_{k}^{A}(\bg; \zeta) &= \bigg[\frac{1}{1-\eps_{+}\eps_{-}}\frac{1-\eps_{-}}{1-\eps_{+}}\frac{1-\zeta^{-1}\eps_{+}}{1-\zeta\eps_{-}}\ \PE\bigg[
-\frac{\bg}{1-\bg}\bigg(-\eps_{+}\eps_{-}+\eps_{-}(1-\zeta)-\eps_{+}(1-\zeta^{-1})\bigg)\bigg]
 \bigg]_{\eps_{+}^{k},\eps_{-}^{k}} \,. 
 \ea
 In fact, this expression implies the very simple relation between the $\Phi$ functions for symmetric and antisymmetric representations
 \be
 \la{dual}
 \Phi_{k}^{A}(\bg; \zeta) = \Phi_{k}^{S}(\bg; \zeta^{-1})\,.
 \ee
 Expanding in $q$ after specialization to the Schur index gives then, \cf (\ref{3.16}), \footnote{See comments after (\ref{3.16}) for the validity of these relations.}
 \bea
 \Phi_{k,0}^{A}(\eta; \zeta) &= \frac{1}{\zeta}-\zeta^{k-1}+\zeta^{k},\\
 \Phi_{k,1}^{A}(\eta; \zeta) &= (\eta+\eta^{-1})\bigg[\frac{1}{\zeta}+\zeta^{k-2}-3\zeta^{k-1}+2\zeta^{k}\bigg]\,, \\
 & \cdots
 \eea
 and so on. Now, due to (\ref{dual}), exponents of $\zeta$ include cases $k-n$ with a non-negative integer $n$. These replace $G_{N}\to G_{N-k+n}$. It makes sense 
 to consider $k\gg 1$ only with $k<N$. We may select here all powers of $\zeta$ with an exponent of that form. 
 In the contour integral
 \ba
 \Phi^{A}_{k}(\bg; \zeta) &= \oint\frac{d\eps_{+}}{2\pi i \eps_{+}^{k+1}}\oint\frac{d\eps_{-}}{2\pi i \eps_{-}^{k+1}} \\
 &\qquad \frac{1}{1-\eps_{+}\eps_{-}}\frac{1-\eps_{-}}{1-\eps_{+}}\frac{1-\zeta^{-1}\eps_{+}}{1-\zeta\eps_{-}}\ \PE\bigg[
-\frac{\bg}{1-\bg}\bigg(-\eps_{+}\eps_{-}+\eps_{-}(1-\zeta)-\eps_{+}(1-\zeta^{-1})\bigg)\bigg]\,, \notag
 \ea
 we pick now minus the pole at $\eps_{+}=1$ and then the pole at $\eps_{-}=1/\zeta$. This defines
 \ba
 \wt \Phi^{A}_{k}(\bg; \zeta) = \zeta^{k-1}(\zeta-1)\, \PE\bigg[\frac{\bg}{1-\bg}\bigg(2-\frac{1}{\zeta}\bigg)\bigg].
 \ea
In conclusion, we can write (up to subleading corrections $\sim q^{N+\delta}$ with $\delta$ being a fixed $k$-independent integer)
\be
R_{A_{k}}^{U(N)}(\eta; q) = 1+[\eta^{N-k}G_{A}^{+}(\eta; q)+\eta^{k-N}G_{A}^{-}(\eta; q)]\,q^{N-k}+\cdots, 
\ee
with
\ba
\frac{G_{A}^{+}(\eta; q)}{G_{\rm D3}^{+}(\eta; q)} &=  \PE\bigg[-\frac{f}{1-f}\bigg](1-\eta q)
\ \PE\bigg[\frac{f}{1-f}(2-\eta q)\bigg] \lp
= (1-\eta q)\ \PE\bigg[\frac{f}{1-f}(1-\eta q)\bigg] =  \frac{1}{1-\eta q}\frac{(\eta q; \frac{q}{\eta})_{\infty}}{(\frac{q}{\eta})_{\infty}}\, .
\ea
Notice also the expression
\be
\la{4.9}
G_{A}^{+}(\eta; q) = -\frac{\eta q}{1-\eta q}\frac{(\frac{q}{\eta})_{\infty}}{(\frac{1}{\eta^{2}}; \frac{q}{\eta})_{\infty}}.
\ee

\subsection{Special limits}

\paragraph{Unrefined limit}
We start with 
\be
\frac{(a q;q)_{\infty}}{(a^{b}q; q)_{\infty}} = \exp\sum_{n\ge 0}[\log(1-aq^{n+1})-\log(1-a^{b}q^{n+1})]=1+(b-1)(a-1)\,\sum_{n\ge 1}\frac{q^{n}}{1-q^{n}}+\mc O((a-1)^{2}),
\ee
Using (\ref{4.9}), we get 
\be
G^{+}_{A}(\eta; q) = -\frac{q}{2(1-q)}\frac{1}{\eta-1}+\frac{q(3q-5)}{4(1-q)^{2}}+\frac{q}{1-q}\sum_{n\ge 1}\frac{q^{n}}{1-q^{n}}+\mc O(\eta-1).
\ee
and this gives
%
the total correction
\be
\la{4.12}
\eta^{N-k}G_{A}^{+}(\eta; q)+\eta^{-N+k}G_{A}^{-}(\eta; q) \stackrel{\eta\to 1}{=} -\frac{q}{1-q}\,(N-k)+\frac{2q}{1-q}\sum_{n\ge 1}\frac{q^{n}}{1-q^{n}}
-\frac{q(2-q)}{(1-q)^{2}}.
\ee
To give an example, for $N=5$ and $k=3$ we get 
\be
1+(\ref{4.12}) \times q^{N-k}=1-4q^{3}-3q^{4}+0\times q^{5}+3q^{6}+\cdots,  
\ee
where we wrote only terms that are not affected by higher giant graviton contributions. This is in agreement with the $\eta=1$ limit of (\ref{2.17}).

\paragraph{$\frac{1}{2}$-BPS limit}

In this limit, we have simply
\be
(1-\eta q)\ \PE\bigg[\frac{f}{1-f}(1-\eta q)\bigg]  = (1-t)\PE[t] = 1\,,
\ee
that implies $G_{A}^{+}(t)=G_{\rm D3}^{+}(t) = -t/(1-t)$, \cf (\ref{3.37}), see also Section 2.4 of \cite{Imamura:2024pgp}.
 
 \section*{Acknowledgements}

We thank 
Alejandro Cabo-Bizet for  discussions related to various aspects of this work. 
Financial support from the INFN grant GAST is acknowledged.

\appendix
\section{Conventions for $q$-functions}
\la{app:special}

We collect in this appendix the definition of  special $q$-functions appearing in the text.

\paragraph{$q$-Pochhammer symbol}

\ba
\la{A.1}
(a; q)_{\infty}&= \prod_{k=0}^{\infty}(1-a\,q^{k})\,, \qquad (a^{\pm}; q)_{\infty} = (a; q)_{\infty}(a^{-1};  q)_{\infty}\, , \\
(q)_{\infty} &\equiv (q; q)_{\infty} = \prod_{k=1}^{\infty}(1-q^{k})\, .
\ea

\paragraph{$q$-theta function}

The $q$-theta function is defined as 
\be
\la{A.3}
\vth(x,q) = -x^{-\frac{1}{2}}(q)_{\infty}(x; q)_{\infty}(qx^{-1}; q)_{\infty}\, ,  
\ee
with 
\ba
\la{A.4}
\vth(x;q) = -\vth(x^{-1};q), \qquad \vth(q^{m}x; q) = (-1)^{m}q^{-\frac{m^{2}}{2}}x^{-m}\vth(x;q)\,.
\ea

\section{A useful differentiation identity}

In our analysis it is useful to give an explicit formula for the double multiple derivative
\be
\partial_{a}^{n}\partial_{b}^{m}e^{X ab}\,.
\ee
It may be obtained by introducing two auxiliary Gaussian variables to have linear dependence on $a,b$ in the exponent:
\ba
& \partial_{a}^{n}\partial_{b}^{m}e^{X ab} = \partial_{a}^{n}\partial_{b}^{m}\frac{1}{\pi}\int dAdB e^{-A^{2}-B^{2}+a\sqrt{X}(A+iB)+b\sqrt X(A-iB)}\lp
= X^{\frac{n+m}{2}}\frac{1}{\pi}\int dAdB (A+iB)^{n}(A-iB)^{m}e^{-A^{2}-B^{2}+a\sqrt{X}(A+iB)+b\sqrt X(A-iB)}\lp
= e^{Xab}X^{\frac{n+m}{2}}\frac{1}{\pi}\int dAdB (A+iB+b\sqrt X)^{n}(A-iB+a\sqrt X)^{m}e^{-A^{2}-B^{2}} \lp
= e^{Xab}X^{\frac{n+m}{2}}\langle  (A+iB+b\sqrt X)^{n}(A-iB+a\sqrt X)^{m}\rangle\, ,
\ea
with 
\be
\langle A^{2p}\rangle= \frac{1}{\sqrt \pi}\Gamma(p+\tfrac{1}{2})\,.
\ee
Introducing complex combinations
\be
z=A+i B, \qquad \bar z=A-iB,
\ee
with the non-zero average
\be
\langle (z\bar z)^{p}\rangle =\frac{1}{\pi} \sum_{q=0}^{p}\Gamma(q+\tfrac{1}{2})\Gamma(p-q+\tfrac{1}{2})\binom{p}{q} = p! \, ,
\ee
gives
\ba
\langle (z+x)^{n}(\bar z+y)^{m}\rangle &= \sum_{a,b}\binom{n}{a}\binom{m}{b}\langle z^{a}\bar z^{b}\rangle x^{n-a}y^{m-b} 
= x^{n}y^{m}\sum_{p\ge 0}p!\binom{n}{p}\binom{m}{p} (xy)^{-p}\, ,
\ea
and thus
\ba
\la{B.7}
\partial_{a}^{n}\partial_{b}^{m}e^{X ab} &= e^{Xab}a^{m}b^{n}X^{n+m} \sum_{p\ge 0}p!\binom{n}{p}\binom{m}{p} (Xab)^{-p}.
\ea
If additional sources linear in $a,b$ are present, the same procedure leads to 
\ba
\la{B.9}
\partial_{a}^{n}\partial_{b}^{m}e^{X ab+\sqrt {X}(\alpha a+\beta b)} &= e^{Xab+\sqrt {X}(\alpha a+\beta b)}X^{\frac{n+m}{2}} 
(\alpha+b\sqrt X)^{n}(\beta+a\sqrt X)^{m}\lp
\sum_{p\ge 0}p!\binom{n}{p}\binom{m}{p} [(\alpha+b\sqrt X)(\beta+a\sqrt X)]^{-p}\, ,
\ea
reducing to (\ref{B.7}) when $\alpha=\beta=0$.

\section{Full details of the main calculation}
\la{app:full}

In this Appendix we present the detailed derivation of (\ref{3.3}, \ref{3.5}, \ref{3.6}). Following \cite{Murthy:2022ien}, we begin by introducing
\ba
\la{C.1}
U&= \sum_{r,s\in\mathbb{Z}+\frac{1}{2}}z^{r}w^{-s}\langle \prod_{n\ge 1}n^{r_{n}+r'_{n}}\partial_{t_{n}^{+}}^{r_{n}}\partial_{t_{n}^{-}}^{r'_{n}}
[\widetilde Z_{\infty}(\bm{t}^{+}, \bm{t}^{-})\, \widetilde K(r,s; \bm{t}^{+}, \bm{t}^{-})]\rangle_{\bm g} \lp
=\frac{\sqrt{zw}}{z-w}\int \prod_{n=1}\frac{dt_{n}^{+} dt_{n}^{-}}{2\pi n g_{n}}\,e^{-\frac{1}{ng_{n}}t^{+}_{n}t^{-}_{n}}
n^{r_{n}+r'_{n}}\partial_{t_{n}^{+}}^{r_{n}}\partial_{t_{n}^{-}}^{r'_{n}}\, 
 \exp\frac{1}{n}\bigg(t_{n}^{+}t_{n}^{-}+t_{n}^{+}(z^{n}-w^{n})-t_{n}^{-}(z^{-n}-w^{-n})\bigg).
\ea
We use (\ref{B.9}) with 
\be
X = \frac{1}{n}, \qquad \alpha = \frac{1}{\sqrt n}(z^{n}-w^{n}) = \frac{1}{\sqrt n}T_{n}^{-}, 
\qquad\beta = -\frac{1}{\sqrt n}(z^{-n}-w^{-n}) = -\frac{1}{\sqrt n}T^{+}_{n}\,,
\ee
and obtain 
\ba
U&= 
\frac{\sqrt{zw}}{z-w}\int \prod_{n=1}\frac{dt_{n}^{+} dt_{n}^{-}}{2\pi n g_{n}}\,
\sum_{p\ge 0}n^{p}p!\binom{r_{n}}{p}\binom{r'_{n}}{p} (t^{+}_{n}-T^{+}_{n})^{r'_{n}-p}(t^{-}_{n}+T^{-}_{n})^{r_{n}-p}\lp
 \exp\frac{1}{n}\bigg(-\frac{1-g_{n}}{g_{n}}t_{n}^{+}t_{n}^{-}+t_{n}^{+}T_{n}^{-}-t_{n}^{-}T_{n}^{+}\bigg).
\ea
To integrate over  $\bm t^{\pm}$, it is convenient to first translate $\bm{t}^{+}$, $\bm{t}^{-}$ according to 
\be
t_{n}^{+}\to  t_{n}^{+}-\frac{g_{n}}{1-g_{n}}T^{+}_{n},\qquad 
t_{n}^{-}\to t_{n}^{-}+\frac{g_{n}}{1-g_{n}}T^{-}_{n}\, .
\ee
This gives
\ba
U&= 
\frac{\sqrt{zw}}{z-w}\int \prod_{n=1}\frac{dt_{n}^{+} dt_{n}^{-}}{2\pi n g_{n}}\,
\sum_{p\ge 0}n^{p}p!\binom{r_{n}}{p}\binom{r'_{n}}{p} 
\bigg(t^{+}_{n}-\frac{1}{1-g_{n}}T^{+}_{n}\bigg)^{r'_{n}-p}\bigg(t^{-}_{n}+\frac{1}{1-g_{n}}T^{-}_{n}\bigg)^{r_{n}-p}\lp
 \exp\frac{1}{n}\bigg(-\frac{1-g_{n}}{g_{n}}t_{n}^{+}t_{n}^{-}-\frac{g_{n}}{1-g_{n}}T^{+}_{n}T^{-}_{n}\bigg).
\ea
We need the diagonal terms in 
\be
\sum_{p\ge 0}n^{p}p!\binom{r_{n}}{p}\binom{r'_{n}}{p} 
\bigg(t^{+}_{n}-\frac{1}{1-g_{n}}T^{+}_{n}\bigg)^{r'_{n}-p}\bigg(t^{-}_{n}+\frac{1}{1-g_{n}}T^{-}_{n}\bigg)^{r_{n}-p}\, ,
\ee
where we will replace
\be
(t^{+}_{n}t^{-}_{n})^{m}\to m! \left(\frac{n g_{n}}{1-g_{n}}\right)^{m}.
\ee
Let us compute the quantity
\ba
& D_{r_{n},r'_{n}}=\sum_{p\ge 0}n^{p}p!\binom{r_{n}}{p}\binom{r'_{n}}{p} 
\bigg(t^{+}_{n}-\frac{1}{1-g_{n}}T^{+}_{n}\bigg)^{r'_{n}-p}\bigg(t^{-}_{n}+\frac{1}{1-g_{n}}T^{-}_{n}\bigg)^{r_{n}-p} \lp
= \sum_{p,a,b\ge 0}n^{p}p!\binom{r_{n}}{p}\binom{r'_{n}}{p} \binom{r'_{n}-p}{a}\binom{r_{n}-p}{b}(t_{n}^{+})^{a}(t_{n}^{-})^{b}
\bigg(-\frac{1}{1-g_{n}}T^{+}_{n}\bigg)^{r'_{n}-p-a}\bigg(\frac{1}{1-g_{n}}T^{-}_{n}\bigg)^{r_{n}-p-b} \lp
=\bigg(-\frac{1}{1-g_{n}}T^{+}_{n}\bigg)^{r'_{n}}\bigg(\frac{1}{1-g_{n}}T^{-}_{n}\bigg)^{r_{n}} \lp
\sum_{p,a\ge 0}n^{p}p!\binom{r_{n}}{p}\binom{r'_{n}}{p} \binom{r_{n}-p}{a}\binom{r'_{n}-p}{a}a! \left(\frac{n g_{n}}{1-g_{n}}\right)^{a}
\bigg(-\frac{1}{(1-g_{n})^{2}}T^{+}_{n}T^{-}_{n}\bigg)^{-p-a} \lp
=\bigg(-\frac{1}{1-g_{n}}T^{+}_{n}\bigg)^{r'_{n}}\bigg(\frac{1}{1-g_{n}}T^{-}_{n}\bigg)^{r_{n}} \lp
\sum_{p,a\ge 0}a!p!\binom{r_{n}}{p}\binom{r'_{n}}{p} \binom{r_{n}-p}{a}\binom{r'_{n}-p}{a} \left(\frac{g_{n}}{1-g_{n}}\right)^{a}
\bigg(-\frac{1}{n(1-g_{n})^{2}}T^{+}_{n}T^{-}_{n}\bigg)^{-p-a}\, .
\ea
The product over $n$ has a factor
\be
\prod_{n\ge 1}(T^{+}_{n})^{r'_{n}}(T^{-}_{n})^{r_{n}}\, .
\ee
The sums of  $\pm$ indices are $\sum_{n}r'_{n}n = |\l'|$ and $\sum_{n}r_{n}n = |\l|$, respectively.
Consider now
\ba
f_{r,r'} &=\sum_{p,a\ge 0}a!p!\binom{r}{p}\binom{r'}{p} \binom{r-p}{a}\binom{r'-p}{a} X^{a}Y^{p} \lp
=\sum_{p,a\ge 0}a!p!\frac{r!}{p!(r-p)!}\frac{r'!}{p!(r'-p)!}\frac{(r-p)!}{a!(r-p-a)!}\frac{(r'-p)!}{a!(r'-p-a)!}X^{a}Y^{p}\lp
=\sum_{p,a\ge 0}\frac{1}{a!p!}\frac{r!r'!}{(r-p-a)!(r'-p-a)!}X^{a}Y^{p}\,.
\ea
Changing summation variable $p+a=q$, we get 
\ba
f_{r,r'}&=\sum_{a,q\ge 0}\frac{1}{a!(q-a)!}\frac{r!r'!}{(r-q)!(r'-q)!}X^{a}Y^{q-a} \lp
= \sum_{q\ge 0}q!\binom{r}{q}\binom{r'}{q}(X+Y)^{q} = {}_{2}F_{0}(-r,-r', X+Y)\,.
\ea
Setting $w=\zeta\, z$, the object we need to evaluate is thus
\ba
\la{C.12}
 \Phi^{S}_{k}&(\bg; \zeta) = \sum_{\l,\l'\vdash k}\frac{1}{\z_{\l}\z_{\l'}}\prod_{n\ge 1}D_{r_{n},r'_{n}} \lp
 = \sum_{\l,\l'\vdash k}\frac{1}{\z_{\l}\z_{\l'}}\prod_{n\ge 1}\bigg(-\frac{1}{1-g_{n}}T^{+}_{n}\bigg)^{r'_{n}}\bigg(\frac{1}{1-g_{n}}T^{-}_{n}\bigg)^{r_{n}} 
  \sum_{p\ge 0}p!\binom{r_{n}}{p}\binom{r'_{n}}{p}\bigg(-\frac{n(1-g_{n})}{T^{+}_{n}T^{-}_{n}}\bigg)^{p}\,,
 \ea
which is a function of $\zeta$ because, after taking the product over $n$, only terms of the form 
\be
T^{+}_{n_{1}^{+}}T^{+}_{n_{2}^{+}}\cdots
T^{-}_{n_{1}^{-}}T^{-}_{n_{2}^{-}}\cdots
\ee
with
\be
n_{1}^{+}+n_{2}^{+}+\cdots = 
n_{1}^{-}+n_{2}^{-}+\cdots, 
\ee
survive, and thus all dependence on $z,w$ is through the ratio  $\zeta=w/z$. The leading giant graviton correction in (\ref{3.3}) is then \footnote{
From $\sum_{s\in\mathbb{Z}+\frac{1}{2}}\zeta^{-s}H_{s} = f(\zeta)$, 
we get  $\mathop{\sum_{N<s}}_{s\in\mathbb{Z}+\frac{1}{2}}H_{s} = \mathop{\sum_{N<s}}_{s\in\mathbb{Z}+\frac{1}{2}}\oint d\zeta\, \zeta^{s-1}f(\zeta)
= \oint d\zeta f(\zeta)\, \sum_{n=0}^{\infty}\zeta^{N+\frac{1}{2}+n-1} = \int d\zeta \zeta^{N-1}\frac{\sqrt{\zeta}}{1-\zeta}f(\zeta) 
= \frac{\sqrt{\zeta}}{1-\zeta}f(\zeta)\bigg|_{\zeta^{-N}}$, and an extra $\frac{\sqrt\zeta}{1-\zeta}$ comes from $\sqrt{wz}/(z-w)$ in (\ref{C.1}).
}
\be
 \sum_{\l,\l'\vdash k}\frac{1}{\z_{\l}\z_{\l'}}\delta Z_{N}^{\l,\l'}(\bg) =Z_{\infty}(\bg)\, \Phi^{S}_{k}(\bg; \zeta) \frac{\zeta}{(1-\zeta)^{2}}
\PE\bigg[-\frac{\bg}{1-\bg}(1-\zeta)(1-\zeta^{-1})\bigg]\bigg|_{\zeta^{-N}}\, ,
\ee
and this proves (\ref{3.4}).
It remains to put in a more explicit form the quantity $\Phi^{S}_{k}(\bg; \zeta)$ which was  defined in (\ref{C.12}).

\subsection{Plethystic representation of $\Phi^{S}_{k}(\texorpdfstring{\bg}{g}; \zeta)$}
\la{app:fullplet}

We start with the following straightforward manipulations
\ba
 \Phi^{S}_{k}(\bg; \zeta) &=  \sum_{\l,\l'\vdash k}\frac{1}{\z_{\l}\z_{\l'}}\prod_{n\ge 1}\bigg(-\frac{T^{+}_{n}}{1-g_{n}}\bigg)^{r'_{n}}
 \bigg(\frac{T^{-}_{n}}{1-g_{n}}\bigg)^{r_{n}} 
  \sum_{p\ge 0}p!\binom{r_{n}}{p}\binom{r'_{n}}{p}\bigg(-\frac{n(1-g_{n})}{T^{+}_{n}T^{-}_{n}}\bigg)^{p} \lp
=\sum_{\l,\l'\vdash k}\frac{1}{\z_{\l}\z_{\l'}}\prod_{n\ge 1} \sum_{p=0}^{\infty}\frac{1}{p!}(-n(1-g_{n}))^{p}\partial_{T^{+}_{n}}^{p}\partial_{T^{-}_{n}}^{p}
\bigg(-\frac{T^{+}_{n}}{1-g_{n}}\bigg)^{r'_{n}}
 \bigg(\frac{T^{-}_{n}}{1-g_{n}}\bigg)^{r_{n}}\lp
=\prod_{n\ge 1} \sum_{p=0}^{\infty}\frac{1}{p!}(-n(1-g_{n}))^{p}\partial_{T^{+}_{n}}^{p}\partial_{T^{-}_{n}}^{p}
\sum_{\l,\l'\vdash k}\frac{1}{\z_{\l}\z_{\l'}}\prod_{n\ge 1}
\bigg(-\frac{T^{+}_{n}}{1-g_{n}}\bigg)^{r'_{n}}
 \bigg(\frac{T^{-}_{n}}{1-g_{n}}\bigg)^{r_{n}}\lp
=\prod_{n\ge 1} \sum_{p=0}^{\infty}\frac{1}{p!}(-n(1-g_{n}))^{p}\partial_{T^{+}_{n}}^{p}\partial_{T^{-}_{n}}^{p} 
\exp\sum_{n\ge 1}\frac{1}{n}\bigg(\eps_{-}^{n}\frac{T^{-}_{n}}{1-g_{n}}\bigg)\bigg|_{\eps_{-}^{k}}
\exp\sum_{n\ge 1}\frac{1}{n}\bigg(-\eps_{+}^{n}\frac{T^{+}_{n}}{1-g_{n}}\bigg)
\bigg|_{\eps_{+}^{k}}\,.
 \ea
 This may be written 
 \ba
 \Phi^{S}_{k}(\bg; \zeta) &= \bigg[\prod_{n\ge 1} \sum_{p=0}^{\infty}\frac{1}{p!}(-n(1-g_{n}))^{p}\partial_{T^{+}_{n}}^{p}\partial_{T^{-}_{n}}^{p} 
\exp\frac{1}{n(1-g_{n})}\bigg(\eps_{-}^{n}T^{-}_{n}-\eps_{+}^{n}T^{+}_{n}\bigg)
\bigg]\bigg|_{\eps_{+}^{k},\eps_{-}^{k}} \lp
=\bigg[\prod_{n\ge 1} \sum_{p=0}^{\infty}\frac{1}{p!}(-n(1-g_{n}))^{p}\bigg(-\frac{\eps_{+}^{n}}{n(1-g_{n})}\frac{\eps_{-}^{n}}{n(1-g_{n})}\bigg)^{p}
\exp\frac{1}{n(1-g_{n})}\bigg(\eps_{-}^{n}T^{-}_{n}-\eps_{+}^{n}T^{+}_{n}\bigg)
\bigg]\bigg|_{\eps_{+}^{k},\eps_{-}^{k}}\lp
=\bigg[\prod_{n\ge 1} 
\exp\frac{1}{n(1-g_{n})}\bigg(\eps_{+}^{n}\eps_{-}^{n}+\eps_{-}^{n}T^{-}_{n}-\eps_{+}^{n}T^{+}_{n}\bigg)
\bigg]\bigg|_{\eps_{+}^{k},\eps_{-}^{k}}\,.
\ea
Adding and subtracting $\frac{1}{1-g_{n}}=1+\frac{g_{n}}{1-g_{n}}$, evaluating the plethystic of the part independent on $g_{n}$, and 
finally rescaling $\eps_{\pm}\to z^{\pm 1}\eps_{\pm}$, we get 
\ba
\Phi^{S}_{k}(\bg; \zeta) &= \bigg[
 \frac{1}{1-\eps_{+}\eps_{-}}\frac{1-\eps_{+}}{1-\eps_{-}}\frac{1-\zeta\eps_{-}}{1-\zeta^{-1}\eps_{+}}\
 \prod_{n\ge 1} 
\exp\frac{g_{n}}{n(1-g_{n})}\bigg(\eps_{+}^{n}\eps_{-}^{n}+\eps_{-}^{n}z^{-n}T^{-}_{n}-\eps_{+}^{n}z^{n}\,T^{+}_{n}\bigg)
 \bigg]_{\eps_{+}^{k},\eps_{-}^{k}}\lp
 = \bigg[
 \frac{1}{1-\eps_{+}\eps_{-}}\frac{1-\eps_{+}}{1-\eps_{-}}\frac{1-\zeta\eps_{-}}{1-\zeta^{-1}\eps_{+}}\
 \prod_{n\ge 1} 
\exp\frac{g_{n}}{n(1-g_{n})}\bigg(\eps_{+}^{n}\eps_{-}^{n}+\eps_{-}^{n}(1-\zeta^{n})-\eps_{+}^{n}(1-\zeta^{-n})\bigg)
 \bigg]_{\eps_{+}^{k},\eps_{-}^{k}} \, ,
 \ea
 which is formula (\ref{3.6}) used in the text. Notice that in free limit $\bg=0$ one has 
 \ba
 \Phi^{S}_{k}(0; \zeta) &=\bigg[
 \frac{1}{1-\eps_{+}\eps_{-}}\frac{1-\eps_{+}}{1-\eps_{-}}\frac{1-\zeta\eps_{-}}{1-\zeta^{-1}\eps_{+}}
 \bigg]_{\eps_{+}^{k},\eps_{-}^{k}} = \frac{1}{\zeta^{k}}-\frac{1}{\zeta^{k-1}}+\zeta\, ,
 \ea
 where we did contour integration, \cf the first line in (\ref{3.16}).

\bibliography{BT-Biblio}
\bibliographystyle{JHEP-v2.9}
\end{document}